 \documentclass[]{emulateapj}
\def\ca{\mbox{[Ca\,{\sc viii}]}}
\def\si{\mbox{[Si\,{\sc vi}]}}
\def\bg{\mbox{Br$\gamma$}}
\def\h2{\mbox{H$_2$}}
\def\fe{\mbox{[Fe\,{\sc ii}]}}
\def\he{\mbox{He\,{\sc i}}}
\def\c2{\mbox{$\chi^2$ }}

\slugcomment{Accepted for publication in the Astrophysical Journal}

\shorttitle{The central parsecs of Centaurus~A}
\shortauthors{Neumayer et al.}

\begin{document}

\title{The central parsecs of Centaurus~A:\\ 
high excitation gas, a molecular disk, and the mass of the black hole\altaffilmark{6}}

\author{N. Neumayer\altaffilmark{1}, M. Cappellari\altaffilmark{2}, J.
Reunanen\altaffilmark{3}, H.-W. Rix\altaffilmark{1},
P.~P. van der Werf\altaffilmark{3}, P.~T. de Zeeuw\altaffilmark{3,4}, R.~I. Davies\altaffilmark{5}}

\altaffiltext{1}{Max-Planck Institute for Astronomy, K\"onigstuhl 17, 69117
  Heidelberg, Germany}
\altaffiltext{2}{Sub-Department of Astrophysics, University of Oxford, Denys Wilkinson Building, Keble Road, Oxford OX1 3RH, England}
\altaffiltext{3}{Leiden Observatory, Leiden University, P.O. Box 9513, NL - 2300 RA 
Leiden, The Netherlands}
\altaffiltext{4}{European Southern Observatory, Karl Schwarzschild-Strasse 2, 85748 Garching bei M\"unchen, Germany}
\altaffiltext{5}{Max-Planck Institute for Extraterrestrial Physics, Postfach 1312, 
85741 Garching, Germany}
\altaffiltext{6}{Based on observations collected at the European Southern Observatory, Paranal, Chile, ESO Programs 74.A-9011(A), 75.B-0490(A)}

%-------------- Abstract -----------------------

\begin{abstract}
We present two-dimensional gas-kinematic maps of the central region in Centaurus~A. The adaptive optics (AO) assisted SINFONI data from the VLT have a resolution of $0\farcs12$ in K-band. The ionized gas species (\bg, \fe, \si) show a rotational pattern that is increasingly overlaid by non-rotational motion for higher excitation lines in direction of Cen~A's radio jet. The emission lines of molecular hydrogen (\h2) show regular rotation and no distortion due to the jet. The molecular gas seems to be well settled in the gravitational potential of the stars and the central supermassive black hole and we thus use it as a tracer to model the mass in the central $\pm 1\farcs5$. These are the first AO integral-field observations on the nucleus of Cen~A, enabling us to study the regularity of the rotation around the black hole, well inside the radius of influence, and to determine the inclination angle of the gas disk in a robust way. The gas kinematics are best modeled through a tilted-ring model that describes the warped gas disk; its mean inclination angle is $\sim 34\degr$ and the mean position angle of the major axis is $\sim 155\degr$. The best-fit black hole mass is $\rm{M_{BH}=(4.5^{+1.7}_{-1.0}) \times 10^7 M_{\odot}}$ (3$\sigma$ error), based on a ``kinematically hot" disk model where the velocity dispersion is included through the Jeans equation. This black hole mass estimate is somewhat lower than, but consistent with the mass values previously derived from ionized gas kinematics. It is also consistent with the stellar dynamical measurement from the same AO observations, which we present in a separate paper. It brings Cen~A in agreement with the $\rm{M_{BH} -\sigma}$ relation.
\end{abstract}


%-------------- Section 1 -----------------------
\section{Introduction}
During the last few years it has been realised that most, if not all, nearby
luminous galaxies host a supermassive black hole (BH) in their nuclei with
masses in the $10^6 - 10^{10} M_{\odot}$ range \citep[e.g.][and references therein]{ferrarese05}. The black hole mass ($\rm{M_{BH}}$) is tightly related with mass or luminosity of the
host stellar spheroid, bulge, (e.g. \citealt{kormendy95,marconi03,haering04}) and with the stellar velocity dispersion, $\sigma$, 
\citep{ferrarese00,gebhardt00}. These correlations have an amazingly low scatter, perhaps surprisingly low, since the quantities $\rm{M_{BH}}$ and $\rm{M_{bulge}}$/$\sigma$ probe very different scales. These facts indicate that the formation of
a massive BH is an essential ingredient in the process of galaxy formation.

 The mass of the black hole at the center of NGC~5128 (Centaurus~A), the most
nearby elliptical galaxy, is still under debate. Centaurus~A hosts a powerful radio source and an AGN revealed by the presence of a powerful radio and X-ray jet (see \citealt{israel98}, for a review; see also \citealt{tingay98,hardcastle03,grandi03,evans04}, and references therein).
Recent stellar dynamical measurements and modeling by \cite{silge05}
result in a black hole mass of 2.4$\times 10^8 M_{\odot}$ (for an edge-on model), while different gas-dynamical studies
found masses in the range of $6.1 \times 10^7 M_{\odot}$
to $2 \times 10^8 M_{\odot}$ depending mainly on the inclination angle of the
modeled gas disk \citep{marconi01,marconi06,neumayer06,krajnovic06}. 
Given a velocity dispersion for NGC~5128 of 138~km\,s$^{-1}$ \citep{silge05}, we would expect a BH mass around $3 \times 10^7 M_{\odot}$ from the $\rm{M_{BH}-\sigma}$ relation. If this BH mass is correct, NGC~5128 has the largest offset from the $\rm{M_{BH}-\sigma}$ relation ever measured (taken the mass values of \cite{marconi01} and \cite{silge05}).
Based on its stellar velocity dispersion and the black hole mass expected by the $\rm{M_{BH}-\sigma}$ relation the radius of influence of Cen~A's black hole is $\sim 0\farcs3$. Thus, ground based, seeing limited observations are not suitable to resolve this radius. 
Using high-spatial resolution ($\sim 0\farcs1$) ionized gas kinematics from adaptive optics assisted and space based observations, \cite{neumayer06} and  \cite{marconi06} find masses reduced by a factor 3-4 compared to previous measurements. 
They were, however, limited to long-slit data at a few position angles and not able to precisely constrain the inclination angle of the modeled gas disk.
With the availability of integral-field spectroscopy (IFS) in the near
infrared (IR) the gas as well as the stars can be mapped in two dimensions
even in dust-shrouded galaxy centers, like Cen~A. Black hole masses can be derived
separately from stars and gas from the same complete data set, and the gas
geometry is constrained by two-dimensional data. The star and gas results can
then be compared to assess the reliability of the modeling techniques. 
SINFONI \citep{eisenhauer03, eisenhauer03b, bonnet04} at the Very Large Telescope (VLT)
combines IFS with the resolving power of adaptive optics assisted observations
and provides data at a spatial resolution of $\sim 0\farcs1$ in K-band. 
For Cen~A the radius of influence of the black hole
at the center should be comfortably resolved. Still, this resolution is far from anywhere near the Schwarzschild radius of the black hole and cannot constrain the central object to be a black hole. However, Cen~A ranks among the best cases for a supermassive black hole in galactic nuclei (see \citealt{marconi06} for a discussion.)

 For our dynamical model we assume a distance to NGC~5128 of 3.5~Mpc to be
consistent with all previous mass determinations. Recent distance measurements
are in the range 3.4~Mpc to 4.2~Mpc \citep{israel98, tonry01, rejkuba04, ferrarese07}, with
typical uncertainties of $\sim8\%$. At the assumed distance of 3.5~Mpc, 1 arcsecond 
corresponds to $\sim$17pc.

 This paper is a follow-on to the work presented in 
\cite{neumayer06} (hereafter HN+06), using high spatial 
resolution Naos-Conica (NaCo) imaging and spectroscopy data to get an accurate 
measurement of Cen~A's black hole mass. While the analysis of HN+06 was restricted to \fe\, 
kinematics along 4 long slit positions, we study in detail the kinematics of different gas species at the center
of NGC~5128 in two dimensions. Unsurprisingly, different gas species exhibit different
behaviors. While the (highly) ionized gas shows (strong) influence by the jet
(both in surface brightness and in the kinematic maps), the molecular gas
(\h2) seems to ``feel" only gravity. This is the reason why
we focus on \h2 when we construct a dynamical model to measure the mass of the
central supermassive black hole. Moreover, the two-dimensional data allow us to determine 
the inclination angle of the gas disk in a robust way. Taken together, these 
advancements over the work of HN+06 and other previous studies, greatly decrease the 
overall uncertainty in Cen~A's black hole mass measurement.
SINFONI stellar kinematics and black hole mass modeling are consistent with the gas-dynamical study
and black hole mass measurement presented here, and will be presented in
a separate paper (Cappellari et al., in prep.).

 The paper is organized as follows:
Section~2 describes the observations and the data reduction. Section~3
presents the gas morphology and kinematics and Section~4 our dynamical
model. Section~5 gives the modeling results and Section~6 discusses them.

%-------------- Section 2 -----------------------

\section{Observations and Data Reduction}

All observations presented here were taken with SINFONI on the UT4 (Yepun) of the VLT of the European Southern 
Observatory (ESO) at Cerro Paranal, Chile, on March 23 and April 1, 2005. SINFONI consists of
a cryogenic near-infrared integral field spectrometer SPIFFI 
\citep{eisenhauer03,eisenhauer03b} coupled to the visible curvature adaptive 
optics (AO) system 
MACAO \citep{bonnet03}. For the Cen~A nucleus, the SINFONI AO module was able to
correct
on an R$\sim$14mag star $36\arcsec$ South-West of the nucleus in excellent seeing of
$0\farcs5$, reaching nearly the diffraction limit of the telescope in the K
band. With the appropriate pixel scale selected
($0\farcs05 \times 0\farcs1$ per spatial element), the spectrograph was able to obtain
spectra across the entire K band (approximately $\lambda$1.93-2.47$\mu$m) at a
spectral resolution of R$\sim$4000 and covering a $3\arcsec \times 3\arcsec $ field of
view, in a single shot. A total of 10 sky (S) and 15 on-source (O) exposures of
900~s each followed a sequence OSO... and were dithered by only a few
pixels to allow removal of bad pixels and cosmic rays. The frames were combined to 
make the final K band data cube with a total on-source exposure of 13500~s.

 In the same way, we obtained H-band data ($\lambda$1.43-1.87$\mu$m) with a
slightly
lower spectral resolution of R$\sim$3000 covering the central $3\arcsec \times 3\arcsec $
with a scale of $0\farcs05 \times 0\farcs1$ per pixel. The overall
integration time for H-band was 3600~s.\\

The data were reduced using the SINFONI data reduction pipeline provided by
ESO. The nearest sky exposure was subtracted from the object frames, 
after which the data was flatfielded and corrected for bad pixels. 
Distortion correction was done based on the position of OH line edges in 
sky frames. The wavelength calibration was based on the nightsky OH 
lines in H-band and on OH lines and arc lamp frames in K-band. Using the 
slitlet positions derived from the sky frames and the obtained wavelength 
calibration a three-dimensional data cube were created from each object 
frame. Finally the slitlet positions were adjusted if necessary based on 
crosscorrelation with NaCo broadband images (taken from HN+06). The typical shift between 
the slitlets is in the range of $\sim 0.1 - 0.3$ pixels. This correction is applied to 
retain the full spatial resolution of the data. After correcting for 
telluric features and flux calibration the data cubes were mosaicked 
together. The stars Hip079775 (B3III) and Hip68359 (B8V) were used both 
as telluric standards and flux calibrators. The flux calibration has an accuracy of about 5-10\%.

%-------------- Section 2.1 -----------------------
\subsection{Spatial resolution}\label{section_psf}

To assess the spatial resolution of the data, the point-spread function (PSF)
is estimated from the unresolved active galactic nucleus (AGN), as discussed
in more detail by HN+06. No additional PSF calibration frames using stars were
taken, since the PSF in adaptive optics observations depends on several
factors such as atmospheric conditions, brightness of and distance to the guide star, contrast
of the guide star to the background, and these conditions are difficult to
duplicate for the science and calibration observations. The most accurate
measure of the science frame PSF in the vicinity of the AGN is therefore
achieved directly on the unresolved nucleus.

 We describe the normalized PSF empirically by a sum of two Gaussian components; one
narrow component describing the
corrected PSF core ($\sigma_{\rm{c}}$) and one broader component
($\sigma_{\rm{s}}$) 
which can be attributed to the seeing halo plus extended emission:
\begin{equation}
PSF(r) = \rm{ \frac{F}{2\pi \sigma^2_{c}}e^{-r^2/2\sigma^2_{c}} +
 \frac{(1-F)}{2\pi \sigma^2_{s}} e^{-r^2/2\sigma^2_{s}}, }
\label{psf_param}
\end{equation}
where F is the ratio of the flux of the narrow component and the total flux of
the PSF (F = $\rm{flux_{c}/flux_{total}}$). The quantity F provides a rough
approximation of the Strehl ratio (S) which gives the quality of an optical
system; S is defined as the observed peak flux divided by the
theoretically expected peak flux of the Airy disk for the optical system
(S = $\rm{peak flux_{c}/peak flux_{Airy}}$).
For the following analysis it is sufficient to measure the quantity F, which
also gives an estimate of the quality of the adaptive optics correction.

\begin{figure}
\epsscale{1.0}
\plotone{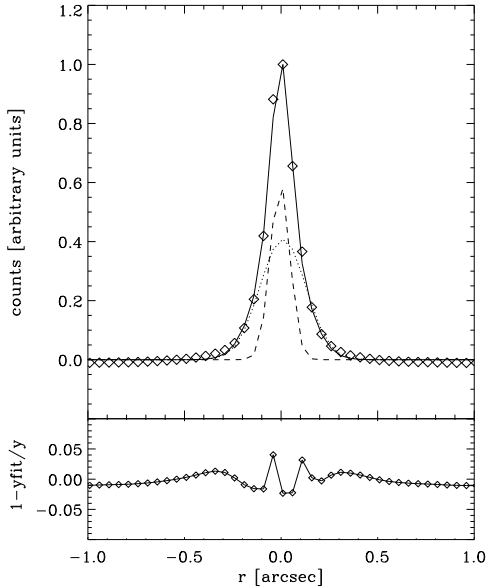}
\caption{Double Gaussian fit of the K-band SINFONI PSF on the continuum just
  next to the \h2 line. The FWHM of the narrow and broad component is  $0\farcs12$ and
  $0\farcs30$, respectively. The estimated Strehl ratio is 17\% (see Section~\ref{section_psf} for a
  definition of these quantities)}
\label{psf}
\end{figure}

 For the K-band cube, a PSF model of the above form is fitted to the peak in
surface brightness in the collapsed wavelength range of $\lambda$2.00 to 2.10
$\mu m$ (just next to the \h2 line). The two components of the PSF model 
shown in comparison to the data in Fig.~\ref{psf} have a full width at half 
maximum (FWHM) of $0\farcs12$ and $0\farcs30$, for the narrow and broad
component
respectively. The estimated Strehl ratio is 17\%. 
The width of the broad component compares well to the FWHM of
the seeing disk as measured by the seeing monitor during the observations 
(FWHM$_V \sim 0\farcs5$ transformed to K-band as in HN+06, FWHM$_K \sim
0\farcs38$).

%-------------- Section 2.2 -----------------------

\subsection{Subtraction of the stellar and non-stellar continua}

\begin{figure*}
\epsscale{1.0}
\plotone{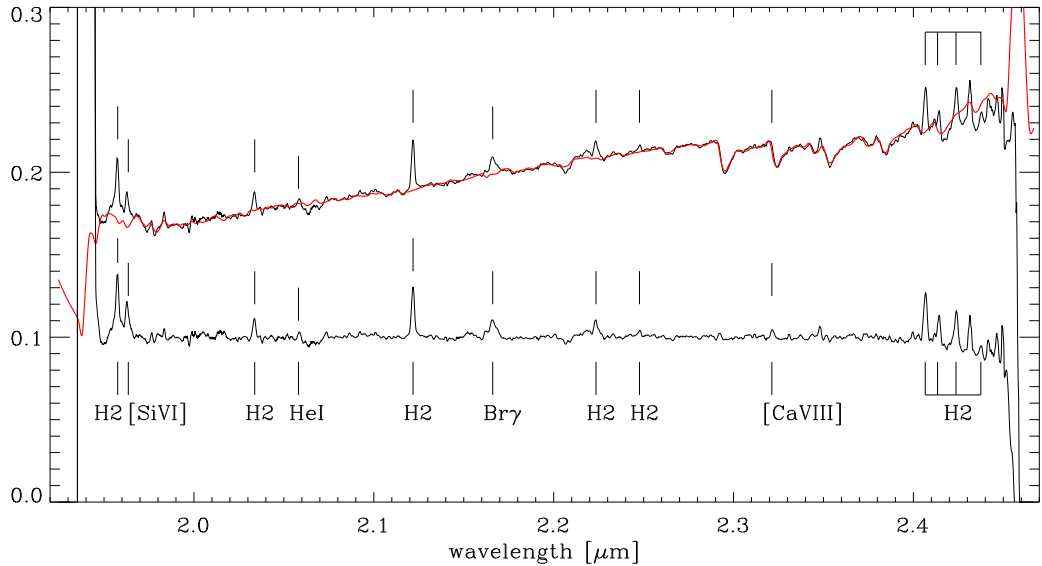}
 \caption{K-band spectrum of the nucleus of Cen~A integrated over a square aperture
 of $0\farcs8 \times 0\farcs8$. The best-fitting composite spectral template 
 (a mix of stellar spectra and a featureless continuum) is
 overplotted in red. The residual, reflecting the pure emission line spectrum, is shown in the
 lower part, with the main emission lines identified. Note the strong CO absorption
 that is very well fitted by the stellar template}
\label{continuum_subtraction}
\end{figure*}

The total or averaged spectrum of the central $0\farcs8 \times 0\farcs8$ of
Cen~A (Fig.~\ref{continuum_subtraction})
shows strong CO absorption lines at 2.3-2.4 $\mu m$ indicating the stellar
continuum.
We use the penalized pixel fitting method (pPXF) of \cite{cappellari04} to fit
the stellar continuum with a positive linear combination of stellar templates. As template stars
serve six late type stars that were observed with SINFONI in the same setup
that we used for the nucleus of Cen~A. This ensures that the
instrumental spectral broadening is the same for the template stars and the
galaxy spectrum, and we need not know the underlying instrumental line
profile.

 As possible template stars we chose the following: K3V, M0III, M0V, M4V, M5II,
and M5III.
The wavelength regions with emission lines were omitted from the fit and the
optimal template was convolved
with Gauss-Hermite expansions \citep{vdmarel93,gerhard93} up to h$_4$,
minimizing the difference between the galaxy spectrum and the template.

In addition to the stellar continuum, the non-stellar continuum is fitted via
an additive Legendre polynomial of fourth order. This becomes important in the
central region where the power-law continuum of the AGN dominates the flux
distribution.
After the subtraction of the stellar and non-stellar continuum we are left with
a data cube of pure emission line spectra.

%-------------- Section 2.3 -----------------------

\subsection{Extraction of the gas emission lines}\label{gas_extraction}

The SINFONI Cen~A spectra in the wavelength range $\lambda$1.43-1.87$\mu$m and
$\lambda$1.93-2.47$\mu$m (the H- and K-band, respectively) exhibit a wealth of
gas emission lines. High excitation lines such as \si\, and \ca, ionized gas 
emission lines (\fe, \bg, \he), and several transitions of molecular hydrogen 
\h2\, are detected (Figure~\ref{continuum_subtraction}). In this paper we focus
on the kinematic properties of \si, \bg, \fe, and \h2\, and show that they
have quite different kinematics. While the high-excitation lines appear to be
affected, or created, by Cen~A's jet, the \h2 gas appears to be solely
rotating. We construct a dynamical model to explain the kinematics of the
strongest line of molecular hydrogen 1-0 S(1) \h2 at $\rm{\lambda=2.122\mu m}$
and use this to measure the mass of the supermassive black hole at the center
of NGC~5128. Single Gaussians provide a good fit to the emission lines and are used to 
measure the central wavelength, width and intensity of each line independently. The fit is performed in 
IDL\footnote{See http://www.rsinc.com} using a non-linear least squares fit to
the line and the errors are the 1-$\sigma$ error estimates of the fit parameters. 
The typical uncertainties of the peak flux and the width of the lines are ~3-5\%, 
whereas the position of the line ca be determined to less than 1\% accuracy. 
This translates to a typical uncertainty in the velocity of the molecular hydrogen line of $\sim \pm$5~km~s$^{-1}$ and in the velocity dispersion of $\sim \pm$10~km~s$^{-1}$.

 To allow for the best possible extraction of the gas lines the
extraction window is centered on the expected wavelength. Its
width is optimized iteratively to fully cover the width of the
line and to make sure the extraction window covers the same range on the
left and right of the line peak. An initial estimate of the Gaussian fit
parameters (amplitude, central position, and width) was derived from a smoothed
spectrum at a central position in the velocity field, near the AGN, and applied 
as a starting value throughout the field. This initial estimate is introduced in order to prevent fitting of spurious lines in the regions of bad pixels or strong continuum variation.

 The signal-to-noise ratio 
of the spectra appears high enough for extracting the gas emission lines and drops rapidly to zero. 
Therefore, we do not spatially bin the data. We only consider the detection of the lines to be secure when their amplitude 
is a factor of 3 above the RMS scatter of the spectrum (A/N$\ge$3).
In this way we get an accurate fit to the lines over the entire field, with typical uncertainties of a few percent.

%-------------- Section 3 -----------------------

\section{Emission-line gas measurements}\label{gas_measurements}

From the parameters of the Gaussian fit - peak value, mean wavelength and
width - we get the flux (or, surface brightness), the velocity and the velocity 
dispersion for the considered gas species, \si, \bg, \fe, and \h2. 
Since the stellar and non-stellar continua have been subtracted, the
line-flux is directly measured as $ \sqrt{2\pi} \rm{F_{peak}} \sigma$. For the
velocity, we take the recession velocity of 
Cen~A's stellar body (v$_{\rm{sys}} =532\pm5$~km~s$^{-1}$, \citealt{marconi01})
as the reference and measure all line shifts with respect to this velocity.

%-------------- Section 3.1 -----------------------

\subsection{Gas kinematics}\label{gas_kinematics}

\begin{figure*}
\plotone{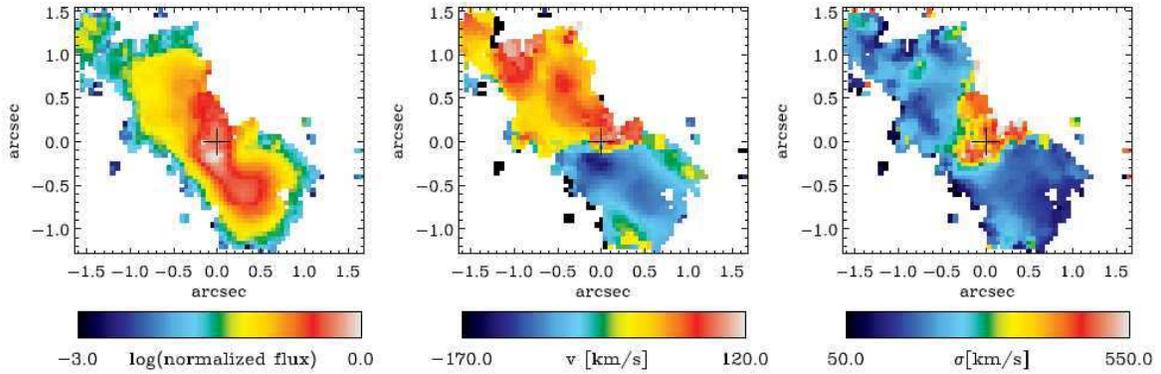}
 \caption{Surface brightness, velocity, and velocity dispersion maps of the \si\, line emission 
 (left, middle, right panel, respectively). The surface brightness map is displayed
 in logarithmic scaling. The velocities and velocity dispersions are only plotted 
 for the regions that are above the lower flux limit displayed in the left panel. The velocity scale is 
 given relative to the systemic velocity of the galaxy (532~km~s$^{-1}$).
 The velocity dispersion is corrected for the instrumental dispersion 
 ($\sigma_{\rm{instr}}\approx 65$~km~s$^{-1}$). North is up and East is to the left}
\label{si}
\end{figure*}

Figures~\ref{si}-\ref{h2} show the maps of total flux, velocity, and velocity 
dispersion for the detected lines of \si, \bg, \fe, and \h2. These line maps illustrate
vividly how the flux distribution and kinematics change when going from high- to 
low-excitation states. The highest excitation line, \si, is dominated by a 
non-rotational component.

 When comparing the velocity fields of \si, \fe, and \h2, (middle panel in
Figures~\ref{si}-\ref{h2}) one notices that the velocity field of \si\, consists
of two major components: rotational and translational motion. The velocity
fields of \fe\, and \bg\, are dominated by rotation but are still distorted by a
non-rotational component that is strongest to the lower right (south-west) of the
field (blue component). These non-rotational motions seen in \si, \fe, and \bg\, are
located close to the projected direction of the radio jet in Centaurus~A 
(P.A.=51$\degr$ \citealt{clarke92}; \citealt{tingay98}; \citealt{hardcastle03}). 
The measured values for the inclination angle of the jet
vary between $50\degr < i < 80\degr$ \citep[][from VLBI data]{tingay98} and
$20\degr < i < 50\degr$ \citep[][from VLA data]{hardcastle03}, but without doubt
the North-Eastern part is pointed towards us.
Surprisingly, the direction of motion seen in \si\, is red-shifted in the jet pointing towards us and
blue-shifted in the counter-jet. Under the plausible assumption that the geometries of the jet 
and \si, are aligned, this indicates an inflow of material towards the nucleus.
 
 This inflow motion could be associated with backflow of gas that was 
accelerated by Cen~A's jet
and after producing a bowshock is flowing back at the side of the jet
cocoon. Although on much larger scales, this phenomenon is seen in jet 
simulations \citep[e.g.][]{krause05}. As the gas is flowing back, it is reionized by radiation from the central source.
This was proposed by \citet{taylor92} as a mechanism to produce the narrow line
regions in Seyfert galaxies.

 While the kinematics of the very high and medium ionization lines \si\, and \fe\, are 
manifestly influenced by this jet induced motion, the velocity field of \h2\,
shows no distortion due to the jet. The velocity field is very smooth and
symmetric and indicative of rotation, with a striking twist of the major kinematic axis around a
median P.A.=$155\degr$. However, the velocity fields in all the other gas tracers give
evidence that at least part of that gas rotates in a nuclear gas
disk. Moreover, the velocity dispersion maps of \si, \bg, \fe, and \h2\, support
the picture of an inclined nuclear gas disk. They all show an elongated structure
in their high dispersion component, mimicking a disk, at a position angle of $125\pm 25 \degr$,
and a declining dispersion profile outwards. The ionized gas species (\si,
\fe, and \bg) show in addition a colder component (with $\sigma \sim 150$
km\,s$^{-1}$) that is elongated in the direction of the translational
motion. Although being as high as $\sigma \sim 400$~km\,s$^{-1}$, the velocity dispersion of \h2 
has the lowest central value. It is not clear what 
causes this high velocity dispersion for the molecular gas (see also \ref{sigma_parametrisation}). 
Nevertheless, \h2 shows the most ordered structure both in the maps of
velocity dispersion and mean velocity and it seems to be well settled in a disk.\\

\begin{figure*}
\plotone{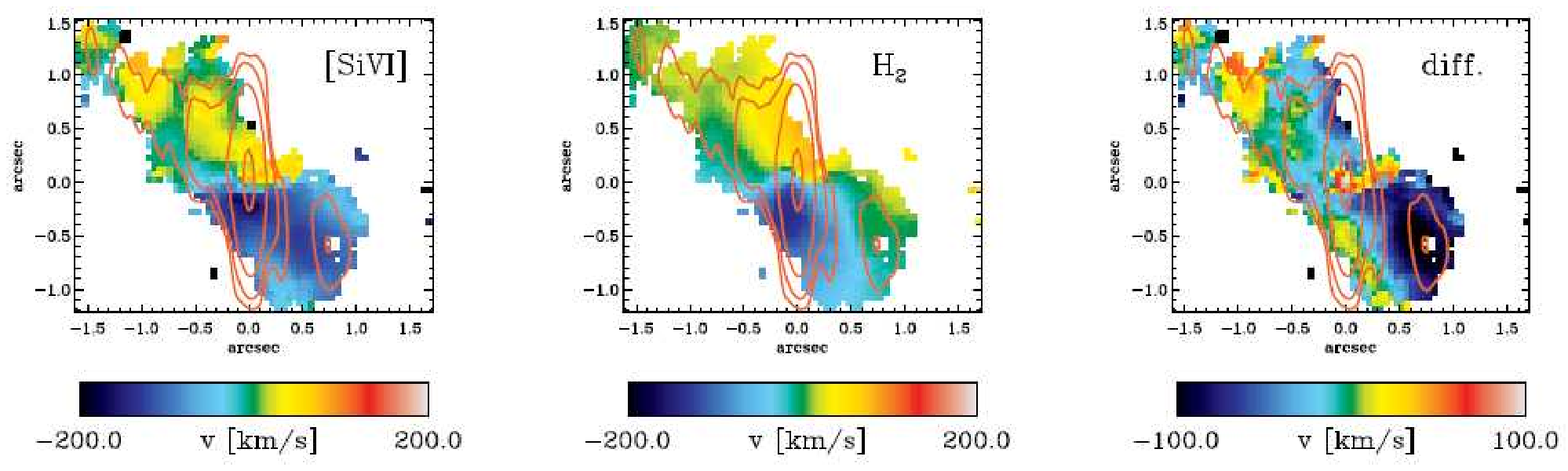}
\caption{Comparison between the \si\, (left) and \h2\, (middle) velocity fields with their difference shown in the right panel. 
The velocity fields are masked with the \si\, flux map which nearly coincides with the radio jet (red contour overlay [unpublished VLA data kindly provided by M. Hardcastle]). The differenced velocity field (right hand panel), meant to `correct' the \si\, kinematics for their rotational component, reveals a distinct \si\, velocity gradient along the jet. The blue component SW of the AGN ($0\farcs7$,$-0\farcs6$) marks the strongest translational velocity component, matching the knot in the radio jet}
\label{compare_si_h2}
\end{figure*}

To visualize the difference in the velocity pattern of \si\, and \h2\, 
Figure~\ref{compare_si_h2} shows the comparison of the \h2\, and \si\, 
velocity fields (left and middle panel) masked with the \si\, flux map. 
The right panel shows the direct difference $v(\si)-v(\h2)$. Overplotted 
are the VLA contours of the radio jet (unpublished VLA data kindly provided by M. Hardcastle). 
The non-rotational component of the \si\, velocity field becomes strongest SW of the 
nucleus, and can be identified with the innermost knot in the VLA radio jet. We
most likely see evidence for jet-gas cloud interaction in the ionized gas species.

%-------------- Section 3.2 -----------------------

\subsection{Gas morphology}\label{gas_morphology}

Looking at the flux maps of the gas species, one notices the very different
morphologies in the high- and low-ionization lines. The morphology of \si\, is
dominated by an elongated structure that extends south-west of the nucleus, at
a position angle of $\sim 33\degr$. This structure widens and ends in a blob
or knot at $\sim 1\arcsec\simeq 17$pc from the center. The position of this knot is
coincident with the innermost knot in the radio counterjet south-west of the
nucleus (seen by \citet{clarke92} and denoted SJ1 by \citet{hardcastle03}).

\begin{figure*}
 \plotone{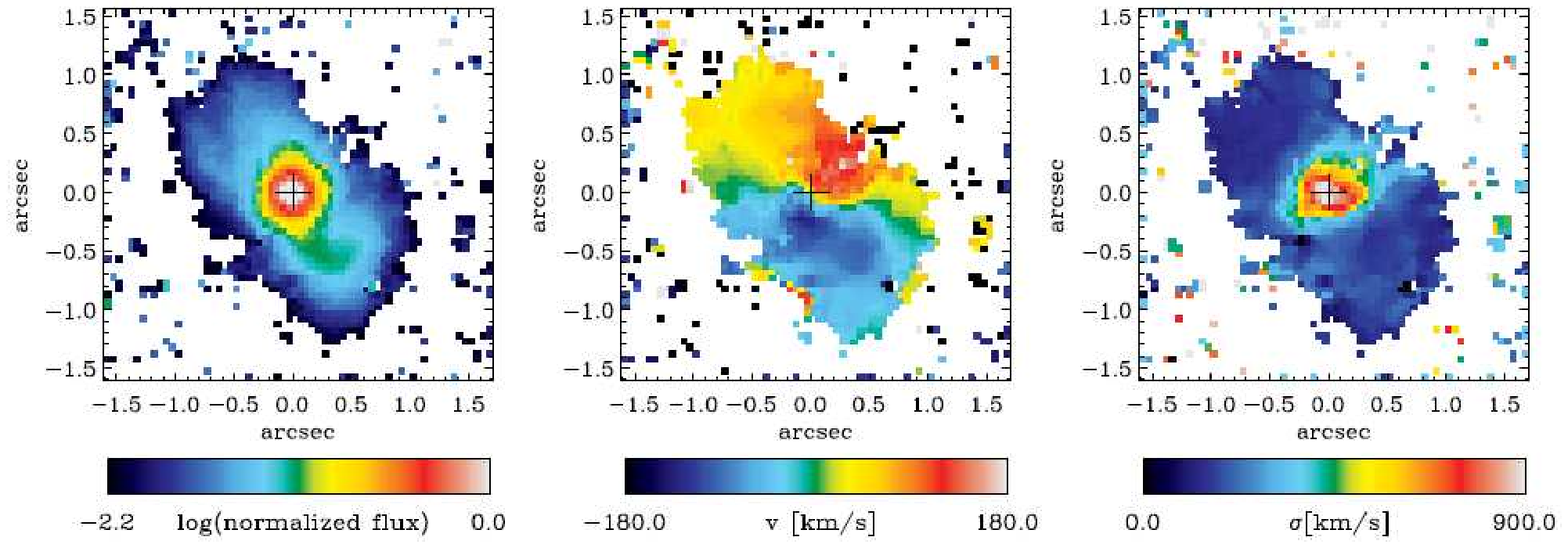}
\caption{Flux, velocity, and velocity dispersion maps of \bg. See caption of 
Fig.~\ref{si} for more details}
\label{bg}
\end{figure*}

\begin{figure*}
\plotone{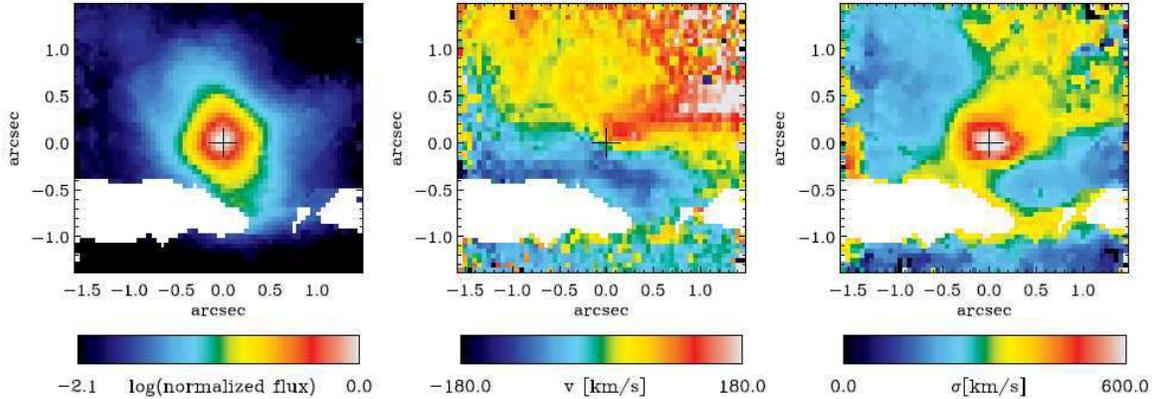}
 \caption{Flux, velocity, and velocity dispersion maps of \fe. See caption of 
 Fig.~\ref{si} for more details. We masked out an artifact around y$\sim -0\farcs7$ that is caused by 
 illumination effects on the SINFONI detector}
\label{fe}
\end{figure*}

 Overall, the morphology of \bg\, and \fe\, (Figures~\ref{bg} and \ref{fe})
resemble that 
of \si\, very closely, but overall the elongation is not as
pronounced and the structure appears rounder.
 
 For all gas species, the P.A. of the elongation is $\sim 33\degr$ which is the
same as for the elongated structure detected in Pa$\alpha$ by
\citet{schreier98}, that is centered on the nucleus and extended by $\pm
2\arcsec$. They interpret this as an inclined, $\sim$40~pc diameter, thin nuclear
disk of ionized gas rather than a jet-gas cloud interaction. However, our 2D data 
show that the gas moves along this elongated structure that
extends around the jet axis. The direction of motion hints towards a backflow of jet 
material onto the accretion disk as mentioned above.

\begin{figure*}
\plotone{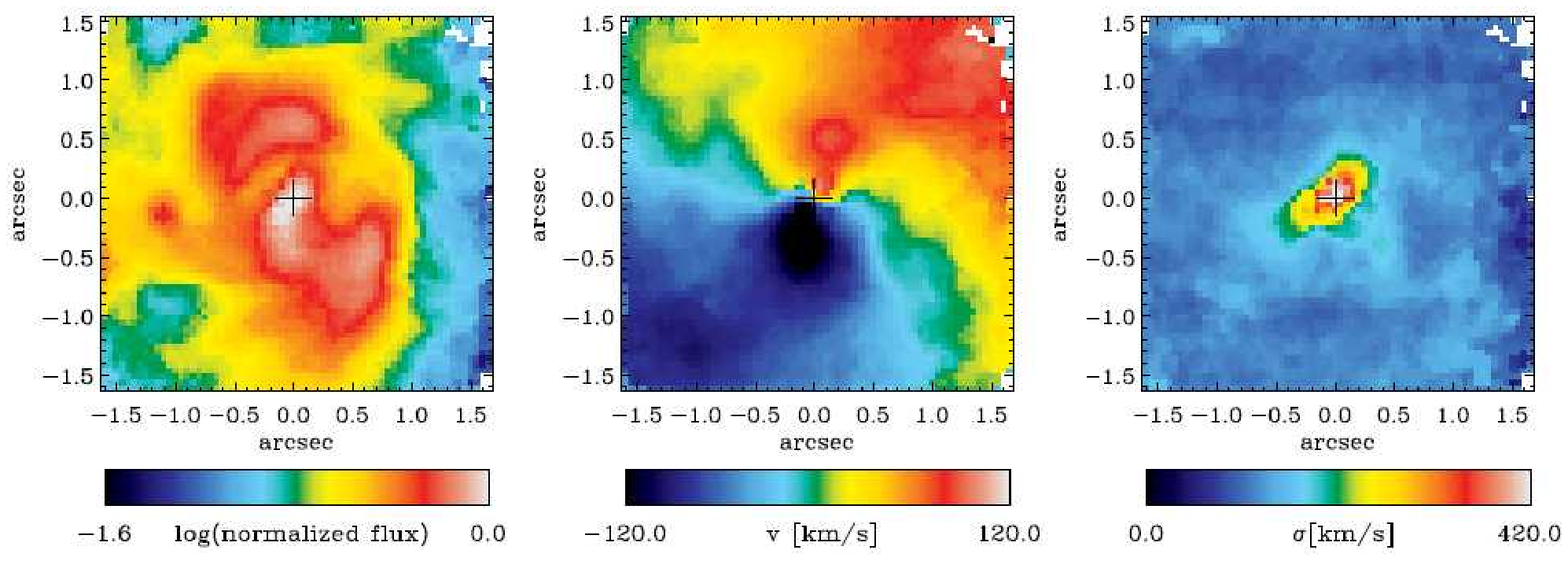}
 \caption{Flux, velocity, and velocity dispersion maps of \h2. See caption of 
Fig.~\ref{si} for more details}
\label{h2}
\end{figure*}

 In addition to this elongated structure, our high spatial resolution integral
field data show evidence for a nuclear disk of ionized and molecular gas
oriented approximately perpendicular to the jet angular momentum vector. Looking at the central
$0\farcs5 \times 0\farcs5$ of the \h2\, flux map (Fig.~\ref{h2}), one notices a 
disk-like structure with a major axis of $\sim
140\degr$. The same structure is visible in the \bg\, and \fe\, flux maps,
although a bit rounder.
Looking at the whole \h2\, flux distribution, it
appears very different from the ionized gas species, and the shells at the
upper left and lower right are a dominant feature. These shells are reminiscent
of
the bowshock structures seen at larger scales in the outer regions of radio
jets \citep[e.g.][for Cyg~A]{carilli88} and on smaller scales in Herbig-Haro
objects \citep[e.g.][]{reipurth02}. They are located at a
(projected) position where the elongated structure in \si, \bg, and \fe\, 
disappears. This is another hint towards the bowshock model of \citet{taylor92}
where the shocked gas (here \si, \fe, and \bg) flows back towards the nuclear source
along the shell. It is important to note that although the shells dominate the appearance 
of the \h2\, flux map, they do not leave any kinematic signature in the \h2\, velocity field.

%-------------- Section 4 -----------------------

\section{Gas Dynamical Modeling\label{modeling}}

While the (highly) ionized gas shows (strong) kinematic influence by the jet, the molecular gas
(\h2) seems to ``feel" only gravity. This is the reason why
we focus on \h2 when we construct a dynamical model to measure the mass of the
central supermassive black hole.

%-------------- Section 4.1 -----------------------
\pagebreak
\subsection{Method}\label{method}

The dynamical model follows the approach of HN+06 and uses two-dimensional 
SINFONI gas kinematic maps as constraints. In brief:
to explain the \h2\, gas motions seen in the center of Cen~A we
construct a kinematic model where we assume the gas moves in a
thin disk solely under the gravitational influence of the
surrounding stars and the expected central black hole. Then, 
the gravitational potential is given as $\Phi(r) =
\Phi_{\star}(r) + \Phi_{\rm{BH}}(r)$. The stellar potential
$\Phi_{\star}$ is taken from
HN+06, where NaCo, NICMOS, and 2MASS K-band
images of NGC~5128 are used to construct a Multi-Gaussian
Expansion (MGE) parameterization to the surface brightness of this
galaxy \citep{emsellem94, cappellari02}. The assumptions of
spherical symmetry and constant stellar mass-to-light ratio
($\rm{M/L_K = (0.72 \pm 0.04)}$ $\rm{M_{\odot}/L_{\odot}}$
\citep{silge05}; HN+06) lead to the three-dimensional mass
model, that gives the stellar velocity contribution to the
dynamical model.

 Our dynamical model is based on the widely used approach to model the emission
line profile of gas moving in a thin disk \citep{macchetto97, vdmarel98,
bertola98, barth01}. In HN+06 we considered three conceptual modeling
approaches: 1) a cold disk model that fully neglects the velocity dispersion, 
2) a hot disk model that accounts for the high velocity dispersion of the gas,
and 3) a spherical Jeans model that accounts for the high velocity dispersion
but neglects the indicated disk geometry. We do not repeat the spherical Jeans
model here, as the \h2\, velocity maps clearly require a disk model.
Since the observed velocity dispersion even of the \h2\, gas at
the nucleus of Cen~A exceeds the mean rotation by more than a factor of
two, we must account for the velocity dispersion in the dynamical model. We
assume the gas disk to be geometrically flat but with an isotropic pressure and
construct an axisymmetric Jeans model in hydrostatic equilibrium. 
In this case, the mean rotation velocity (azimuthal velocity)
$\overline{v_{\Phi}}$ is given by the Jeans equation \citep[][Eq.4-64a]{BT87}
\begin{equation}
\overline{v_{\Phi}}^2 = \frac{R}{\rho_g} \frac{\partial(\rho_g
\sigma_R^2)}{\partial R} + R
\frac{\partial\Phi}{\partial R},
\label{azimuthal_vel}
\end{equation} 

where $R$ is the projected radius and $\sigma_R$ is
the radial velocity dispersion of the gas. We assume that the \h2 surface
brightness of the gas disk $\Sigma_g$ reflects the tracer gas density
$\rho_g$ (see Section~\ref{isbd_parametrisation} for a discussion).
Note that the model is not self-consistent, i.e. the contribution of the gas mass
to the overall potential is neglected. We estimate the mass inside the SINFONI 
field-of-view, following \cite{israel90}. They
identify a central \h2\, gas disk with an outer radius of $\sim 6\farcs5$, 
a thickness of $\sim 3\farcs3$, and an inner cavity of $\sim 1\farcs7$
(assuming a distance to Cen~A of D=3.5\,Mpc), and with a density distribution of 
$\rho(r) \sim r^{-2}$. The total disk mass is $2\times 10^7M_{\odot}$. 
To get a first order estimate of the mass inside the central $3\arcsec \times3 \arcsec$ 
we assume a constant density throughout the disk and neglect
the inner cavity, which is not confirmed by our data. Inside 
the field of view of our SINFONI observations we therefore expect 
$M_{\h2}\sim 1\times 10^{6}M_{\odot}$, which is a factor of 60 to 200 smaller 
than the black hole mass measurements for Cen~A (HN+06; \citealt{marconi01}), 
and therefore negligible.

 To match the observations, the
resulting velocity field can then be projected onto the plane of
the sky, given the inclination angle $i$ and the position angle (P.A.) of
the projected major axis of the gas disk.  
For comparison with the data, the projected velocity field must also be
broadened
with the velocity dispersion and then weighted by the gas surface
brightness, using the parameterizations given subsequently in Sections
\ref{sigma_parametrisation} and
\ref{isbd_parametrisation}, respectively. Finally, we simulate observations of the disk through the SINFONI instrument,
i.e. we convolve with the SINFONI PSF and sample over the pixel size, to
achieve the
best possible match to the data. The model thus has three free parameters, the
black hole mass $\rm{M_{BH}}$, the inclination angle of the disk $i$, and its
projected P.A. $\zeta$. The galaxy center position on the detector 
(defined as the continuum peak) and the
stellar mass-to-light ratio are fixed beforehand, with 
$\rm{M/L_K = 0.72}$ $\rm{M_{\odot}/L_{\odot}}$ as described above. 
To measure the black hole mass
in NGC~5128, we run models with different values for the free parameter and
look for the best possible match to the data, minimizing the \c2.

 For the modeling we use the IDL software\footnote{using Craig B. Markwardt's
MPFIT package}
of HN+06, which accounts for the SINFONI PSF (as determined in
Section \ref{section_psf}), instrumental broadening, and the finite SINFONI
pixel
size, to generate a two-dimensional model spectrum with the same pixel scale as
the
observations. The extraction of the mean velocity and velocity
dispersion from this synthetic data cube is carried out in exactly the same
manner as for the observed data, by
fitting single Gaussians to the individual spectra at each pixel.

%-------------- Section 4.2 -----------------------

\subsection{Intrinsic gas velocity dispersion}\label{sigma_parametrisation}

The observed velocity dispersion of the molecular hydrogen
\h2\, gas at the center of NGC~5128 peaks at 400~km~s$^{-1}$ and
exceeds the measured mean rotational velocity by more than a factor of two 
(for comparison: the ionized gas species \fe\, used in the study of HN+06, has a 
peak velocity dispersion of $\sim$ 600~km~s$^{-1}$).
The physical origin of this high velocity dispersion is not clear.
Partially it might be explained by spatially unresolved rotation
\citep{marconi01, marconi06}, or it might be due to local
turbulent gas motions, as suggested by several authors for other
galaxies \citep[e.g.][]{barth01, vdmarel98, verdoes02}, which appears problematic
especially for the \h2.

 Regardless of its physical cause, we consider the high velocity dispersion to
contribute to the pressure
support of the gas disk. Following the approach of
HN+06 we include the velocity dispersion to the gas
dynamical model via an isotropic pressure term in the Jeans
equation. The rotational velocity therefore becomes sub-Keplerian.
% and matches the observed data better than a cold disk model. 
We
find that the intrinsic velocity dispersion is well described by a
double exponential profile of the form
\begin{equation}
\sigma_{R} = \sigma_{0} e^{-r/r_0} + \sigma_{1} e^{-r/r_1},
\label{sig_param}
\end{equation}
and we fit the
observed \h2, velocity dispersion profile for the best set of parameters
($\sigma_{0}$=140~km~s$^{-1}$, $ \sigma_{1}=65$~km~s$^{-1}$, $r_0=0\farcs25$, and
$r_1=4\farcs0$) to get the intrinsic dispersion profile.

%-------------- Section 4.3 -----------------------

\subsection{Emission line surface brightness}\label{isbd_parametrisation}

For various aspects of the modeling, we need to know the intrinsic spatial
emission
line profile (assumed to be axisymmetric). But a direct deconvolution of the
observed
surface brightness is very difficult, since for adaptive optics observations 
the exact shape of the PSF is unknown and we would therefore introduce artifacts
to the
gas distribution that bias
the velocity distribution inside the inner $0\farcs5$.

 Furthermore, the gas morphology of \h2\, is quite complex and it is
not a priori clear how this complex morphology leads to such a
smooth velocity field. On the other hand, the surface brightness does not
necessarily resemble the real physical gas structure. The bulk mass in \h2 is most
probably at very low temperatures (T$\sim 10-15$~K, as derived by \citealt{israel90} 
from CO observations) and therefore not excited.

\begin{figure*}
\epsscale{0.8}
\plotone{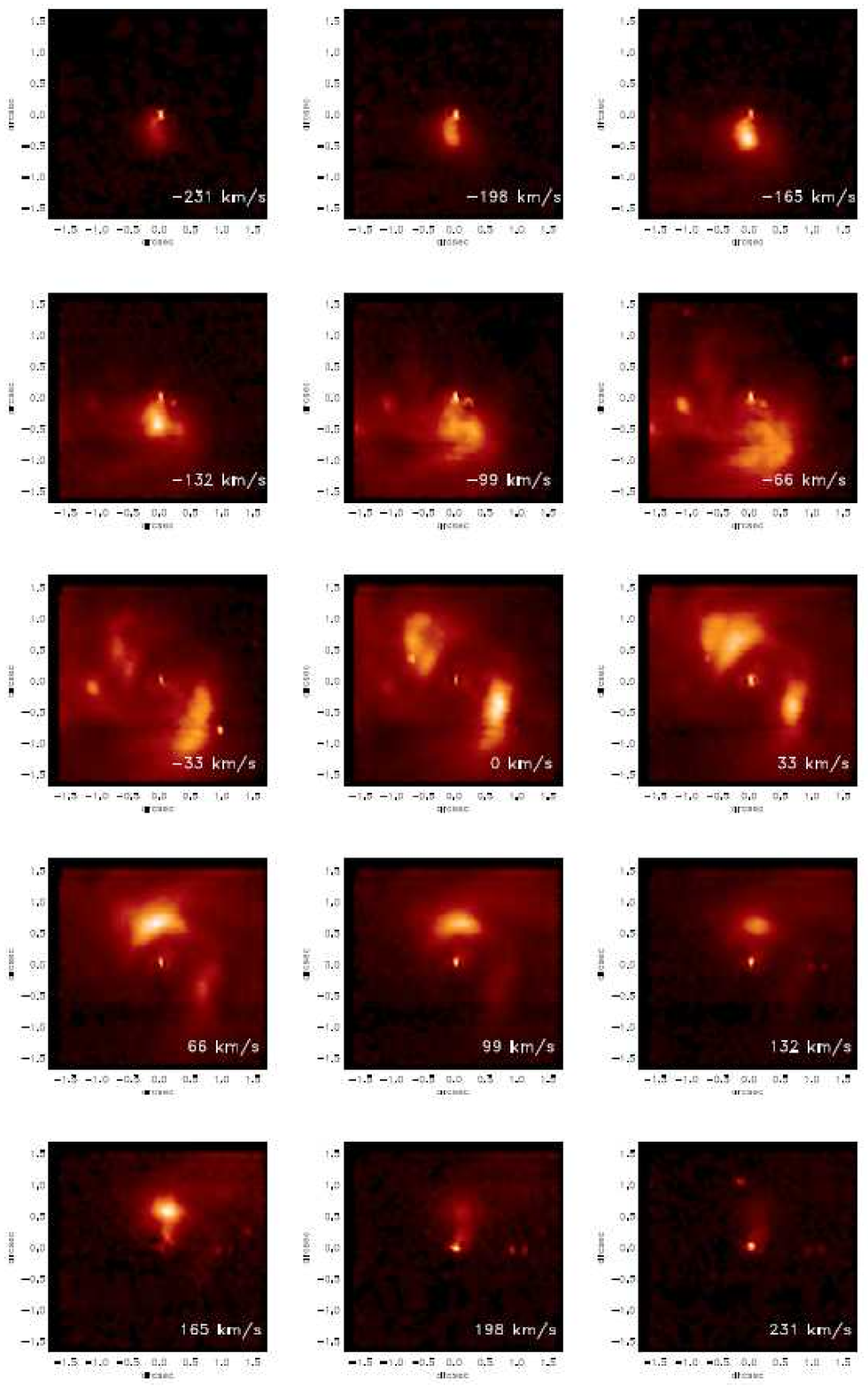}
\caption{Velocity channel maps for the molecular gas (\h2\,) at $\lambda$2.121~$\mu$m observed in the K-band. The
  middle panel shows the flux in the velocity bin that corresponds to the
  recession velocity of the overall galaxy (532~km~s$^{-1}$). The other bins
  slice the cube at blue- and redshifted velocities to this zero position. The width of the velocity slices
  is 33~km~s$^{-1}$}
\label{slice_h2}

\end{figure*}

 Figure~\ref{slice_h2} shows slices through the 
data cube around the \h2 line. The width of the velocity slices is 33~km~s$^{-1}$, corresponding to one pixel in wavelength direction (i.e. half the spectral resolution $\sigma_{\rm{instr}}$). Material located near the nucleus is present in several consecutive panels, indicating that the material has a high velocity dispersion. In addition
there is material that appears in shell-like structures only in one or two
bins, hinting towards a lower velocity dispersion.
The surface brightness of the \h2\, gas seems to be highest along the jet 
direction for $0\farcs5 < \rm{r} < 1\farcs3$ (see Figure~\ref{h2}). 
We actually see shell-like structures that might be due to shocked \h2 gas; 
this hypothesis is supported by the fact that the \si\, and \bg\, gas
distributions 
fit quite nicely into the shells seen in \h2.

Looking at the flux distribution of \h2\, (Figure~\ref{h2}), in the radial range
$\rm{r} < 0\farcs4$ a elongated structure which might be reminiscent 
of a disk is visible with a major axis of P.A.=$136\degr$. Ellipse fits to this
disk structure give
a minor-to-major axis ratio of ${\rm q}\sim 0.67$ which translates to an
inclination angle of
$\sim 48\degr$ given a circular thin disk configuration. The hypothesis of
the central disk structure is also supported by the shape of the
velocity dispersion field. Looking at the right panel of
Figure~\ref{h2} we see a disk-like structure with a
major-axis position angle of $\sim 140 \degr$ inside $\sim 0\farcs4$.

 Drawing on this qualitative description of the \h2\, line
distribution, and the conjecture that the gas moves in a disk, 
we model the emission line surface brightness in two parts: an
exponential disk that dominates the inner
$0\farcs5$ and a smoothed version of the actually observed \h2\, flux distribution 
resembling the detailed gas morphology in the region outside $0\farcs5$, where
the PSF convolution is less critical.
For the black hole mass modeling it is not important to add the second
(larger) component, since the inner $<0\farcs5$ are dominant. Nevertheless we
add it
to get a better estimate of the influence of the non-symmetric gas distribution
on the appearance of the gas velocity field, since part of the small-scale 
structure in the velocity fields can be reproduced by a
patchy gas surface brightness folded into the model \citep{barth01}.

 We have tested the influence of the parameterization of the exponential disk
component on the resulting best-fit black hole mass, and find that a profile
that puts $\sim$12\% more (less) flux inside the central $0\farcs5$ 
(but still fits the observed flux distribution well) results in a black hole
mass that is less than 3\% lower (higher). This result is in line with the 
extensive tests on the influence of the surface brightness parameterization 
carried out by \cite{marconi06}. In any case, this uncertainty is small 
compared to other uncertainties that enter the black hole mass measurement, 
e.g. the inclination angle of the gas disk.

%-------------- Section 4.4 -----------------------

\subsection{Tilted-ring model}\label{kinemetry_Section}

The kinematics predicted by a flat, or co-planar, thin-disk model must be an
oversimplification, as becomes apparent when comparing to the observed data. 
The twists in the velocity field cannot be
reproduced, and the velocity gradient is strictly declining from the peak at
$\sim 0\farcs2$ outwards without being able to resemble the second and third
peak at $0\farcs45$ and $1\farcs 1$ ($\sim 19$pc). 
We therefore model the kinematics via a
tilted-ring model \citep{begeman87} as it was done before on scales of $r\sim 0\farcm5$ to
$3\arcmin$ by
\cite{quillen92} and \cite{nicholson92} who modeled the CO(2-1) and
H$\alpha$ velocity fields in Cen~A, respectively.
The difference to a coplanar model is that the inclination
angle and position angle of the gas disk are a function of radius.
The orbits of the gas at each radius remain circular, but neighboring
orbits are not necessarily in the same plane. The gas-disk
geometry changes from co-planar to warped.

\begin{figure}
%\epsscale{0.7}
\plotone{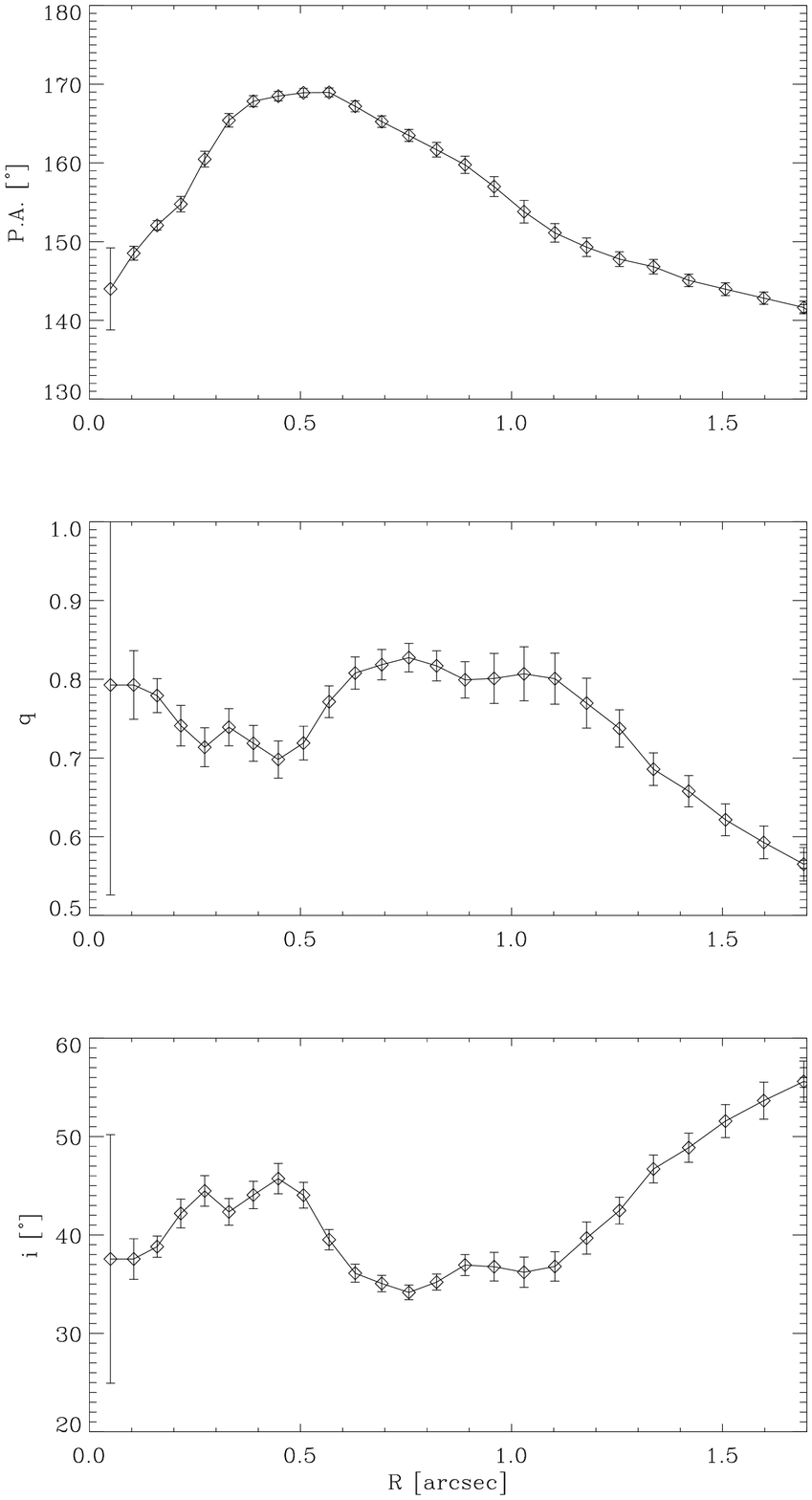}
 \caption{Disk parameters of the kinemetry fit to the velocity field as a function of projected
radius. The inclination angle $i$ is derived from the flattening $q$
($i=\arccos(q)$)}
\label{kinemetry_major}
\end{figure}

\begin{figure*}
\epsscale{0.9}
\plotone{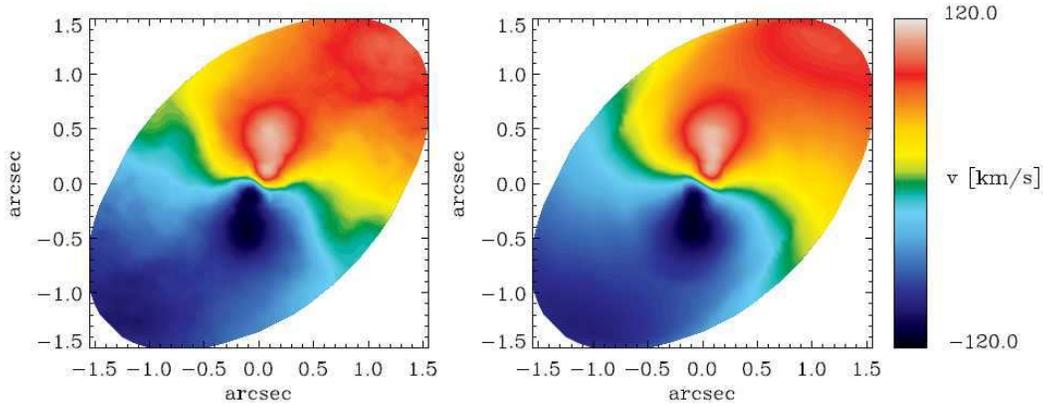}
 \caption{The symmetrized \h2 velocity field (left panel) compared to the best fit kinemetry model (right panel).
 See Section~\ref{kinemetry_Section} for a description of the kinemetry model. The ellipse shows the flattening of the largest circle}
\label{kinemetry_map}
\end{figure*}

 We work in polar coordinates and use discrete radial steps where the model is
to be calculated. The model is linearly interpolated between the discrete
points on the model grid. The gas disk is made up of concentric rings. Each
ring is represented by three parameters: its radius, $R$, inclination angle
$i$, and azimuthal angle $\zeta$ (relative to the projected major axis). When
projected
along the line-of-sight, the rings become ellipses. The flattening, $q$, of the
major and minor axes is related to the inclination angle $i$ via $\cos(i) =
q$. The flattening defines an ellipse on the sky of ellipticity $\epsilon =
1-q$. If the gas is assumed to move on circular orbits along the rings, the
projected velocity can be described by the simple cosine form
\begin{equation}
 v(R,\zeta) = v_{sys} + v_c(R) \sin(i(R)) \cos(\zeta(R)),
\label{vel_field}
\end{equation}
where $v_c(R)$ denotes the circular velocity on a given ring of radius $R$, while
$v_{sys}$ gives the systemic velocity of the entire galaxy.
This method of expanding the full gas velocity field in a set of tilted rings
goes back to the work of \cite{begeman87}. Here, we are using the method of 
\cite{davor06}, called kinemetry, to determine the set of ellipses
along which the velocity field is best described by equation~(\ref{vel_field}). 
This method is a generalisation of surface photometry to all moments of the 
line-of-sight velocity distribution. It performs harmonic expansion of 2D maps 
of observed kinematic moments (velocity, velocity dispersion, and higher Gauss-Hermite terms) along 
the best fitting ellipses (either fixed or free to change along the radii). The
parameters of the best-fit kinemetry model, using 27 tilted rings, are plotted 
in Figure~\ref{kinemetry_major} and
listed in Table~\ref{kinemetry_params}, while Figure~\ref{kinemetry_map} shows
the fitted circular velocity map, using these parameters. Note that the
parameters of the tilted-ring model vary smoothly with radius, although we do
not restrict the kinemetry routine to smooth functions $i(R)$ and $\rm{PA}(R)$. 
We refer the interested reader to \cite{davor06} for a detailed description of the kinemetry method.

\begin{table}[h]
\begin{center}
\caption{Parameters of the tilted ring model\label{kinemetry_params}}
\vspace{0.2cm}
\begin{tabular}{l c c c c c}
\tableline\tableline
R [$\arcsec$]& P.A. [$\degr$] & $ q$ & & i [$\degr$]& $v$ [km \,s$^{-1}$]  \\
\tableline

0.05  &   144.0  &  0.71   & &   45.0   &  55. \\                     
0.10  &   148.5  &  0.77   & &   37.6   & 106.  \\                     
0.16  &   152.1  &  0.79   & &   38.8   & 112.  \\                     
0.22  &   154.8  &  0.70   & &   42.2   & 107.  \\                     
0.27  &   160.5  &  0.66   & &   44.5   & 110.  \\                     
0.33  &   165.4  &  0.56   & &   42.3   & 110.  \\                     
0.39  &   167.8  &  0.60   & &   44.1   & 113.  \\                     
0.45  &   168.5  &  0.64   & &   45.7   & 115.  \\                     
0.51  &   168.9  &  0.67   & &   44.0   & 109.  \\                     
0.57  &   169.0  &  0.78   & &   39.5   & 103.  \\                     
0.63  &   167.2  &  0.81   & &   36.1   &  96.  \\                     
0.69  &   165.2  &  0.82   & &   35.1   &  89.  \\                     
0.76  &   163.5  &  0.82   & &   34.2   &  82.  \\                     
0.82  &   161.7  &  0.82   & &   35.2   &  79.  \\                     
0.89  &   159.8  &  0.80   & &   36.9   &  76.  \\                     
0.96  &   157.0  &  0.80   & &   36.8   &  76.  \\                     
1.03  &   153.8  &  0.80   & &   36.2   &  75.   \\                     
1.10  &   151.1  &  0.80   & &   36.8   &  75.   \\                     
1.18  &   149.3  &  0.76   & &   39.7   &  76.   \\                     
1.26  &   147.8  &  0.74   & &   42.5   &  77.   \\                     
1.34  &   146.8  &  0.69   & &   46.7   &  80.   \\                     
1.42  &   145.1  &  0.65   & &   48.9   &  83.   \\                     
1.51  &   144.0  &  0.61   & &   51.6   &  85.   \\                     
1.60  &   142.8  &  0.58   & &   53.6   &  87.  \\                     
1.69  &   141.6  &  0.56   & &   55.6   &  88.   \\                     
1.79  &   137.9  &  0.64   & &   54.6   &  90.   \\                     
1.90  &   134.9  &  0.58   & &   58.9   &  89.   \\                     

\tableline\tableline  
\end{tabular}
\end{center}
\end{table}

%-------------- Section 5 -----------------------
 
\section{Results}
We now focus on modeling the \h2\, gas kinematics in the central $1\arcsec$
($\sim 17$pc) of NGC~5128 with a warped disk model where the gas orbits 
in the joint potential of the stars and the central black hole.
The central black hole mass as well as the overall inclination angle 
of the gas disk are varied to find the 
best-fit to the observed velocity and velocity dispersion maps.
 
%-------------- Section 5.1 -----------------------
 
\subsection{Structure of the \h2 disk}\label{h2_structure}

The rotational velocity field seen in molecular hydrogen shows beautiful
symmetry about the center despite the twist that alters the projected
orientation of the
kinematic major axis by $\pm 14\degr$ about a mean value of $\sim
155\degr$ (see Table~\ref{kinemetry_params}). Especially for the central 
values, the orientation of the \h2 gas disk is consistent with an orthogonal 
disk-jet picture, as the radio jet in Cen~A is located at a position angle of 
$51\degr$ \citep{tingay98}.

 In addition to the twist in position angle there is a superimposed variation
of inclination angle that causes the maximum of the projected rotational
velocity to vary by $\pm$ 30~km~s$^{-1}$ (around a mean value of 90~km~s$^{-1}$, cf. Table~\ref{kinemetry_params}). This variation is indeed significant compared to the uncertainty of the velocity which is $\sim \pm$5~km~s$^{-1}$ (see Section~\ref{gas_extraction}.

\begin{figure*}
\epsscale{0.9}
\plotone{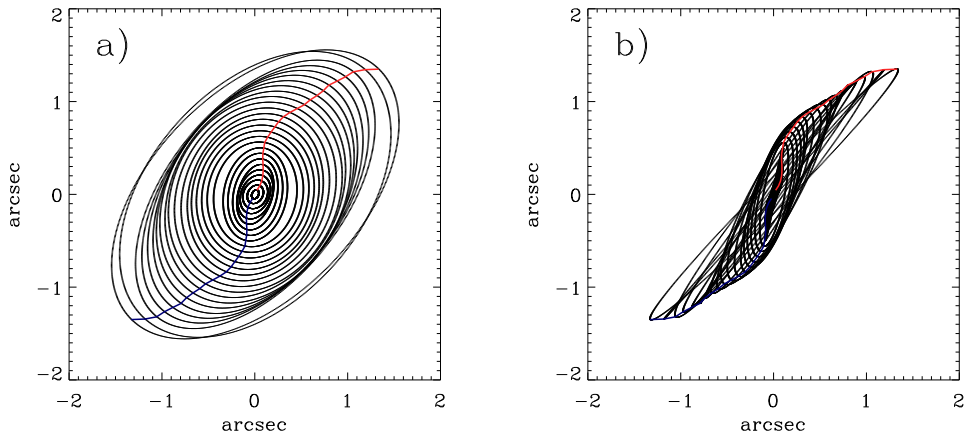}
 \caption{Tilted-ring model to describe the nuclear \h2, gas disk seen at (a) the original 
 orientation as fitted by kinemetry and (b) rotated by $45\degr$ to make the
 warp more visible. The red and blue lines are the line-of-nodes for 
 the receding and approaching side, respectively}
\label{disk_structure}
\end{figure*}

 The structure of the gas disk as fitted by the tilted-ring model is 
presented in Figure~\ref{disk_structure}. For this fit the velocity field was symmetrized 
using the assumption $v(x,y)=-v(-x,-y)$. The left panel (a) shows the actual kinemetry
fit to the symmetrized velocity field. In panel (b) the disk is rotated by 45$\degr$ to an almost edge-on 
view, to make the warp clearly visible. The position angle twist is indicated by the 
red and blue lines for the receding and approaching side, respectively.

Our SINFONI integral field data allow us to constrain the inclination angle of
the gas disk very well. Its mean value is $45\degr \pm 7\degr$, which is in very good
agreement with the best-fitting inclination angle measured by HN+06 from their 
NaCo long-slit data ($45\degr$). The inclination angle is also in agreement with 
the value derived by \cite{hardcastle03} from VLA data ($20\degr < i < 50\degr$), 
it is however somewhat smaller than the value from VLBI data derived by 
\cite{tingay98} ($50\degr < i < 80\degr$).

 The central \h2\, gas kinematics are well described by the 
tilted-ring model. As already stated above, a coplanar disk model would not be able to
reproduce the twist in the rotational velocity field and would therefore
represent an oversimplification of the kinematic structure. The coplanar disk model
can be ruled out at almost 3$\sigma$ confidence level.
 
 The warp of the larger scale disk ($>10\arcsec$) was probably caused by the merger
event, that
occurred a few times $10^8$ years ago \citep{quillen93, peng02}. It
is not a priori clear whether the warp in the innermost arcsecond is connected
to this larger scale warp that creates the prominent appearance of the dust
disk in Cen~A \citep{quillen06}. For the nuclear gas disk on scales $\la 1\farcs2$ ($\la 20$pc)
self-induced warping of the accretion disk \citep{pringle96} might be the
driving mechanism. 
Our model is not intended to explain the origin of the warp, but
is rather meant as a geometric model to resemble the observed
velocity field. 

%-------------- Section 5.2 -----------------------
 
\subsection{Importance of the inclination angle}

The best-fit black hole mass in gas dynamical models depends strongly on
the inclination angle assumed for the gas disk model ($\rm{ M_{BH}} \varpropto
\sin(i) $). Any uncertainty in the disk inclination angle will directly propagate to the
uncertainty in the black hole mass.

 Our tilted-ring model fitted by kinemetry fixes the overall shape of the gas disk in 
the model. However, we allow for overall variations in the inclination angle, not to
overconstrain the model. That means the position angles for the tilted-rings remain fixed, 
as well as their relative orientation. While we fix the overall `twist' of the disk a priori, 
we allow the full dynamical model to re-fit the global inclination.
In that way we make use of the geometrical information that is contained in the velocity
field (using kinemetry), but at the same time give our dynamical model the freedom to
find the best inclination angle given the general assumptions of the model.

%-------------- Section 5.3 -----------------------
 
\subsection{Best-fit model and black hole mass}

\begin{figure*}
\epsscale{0.9}
\plotone{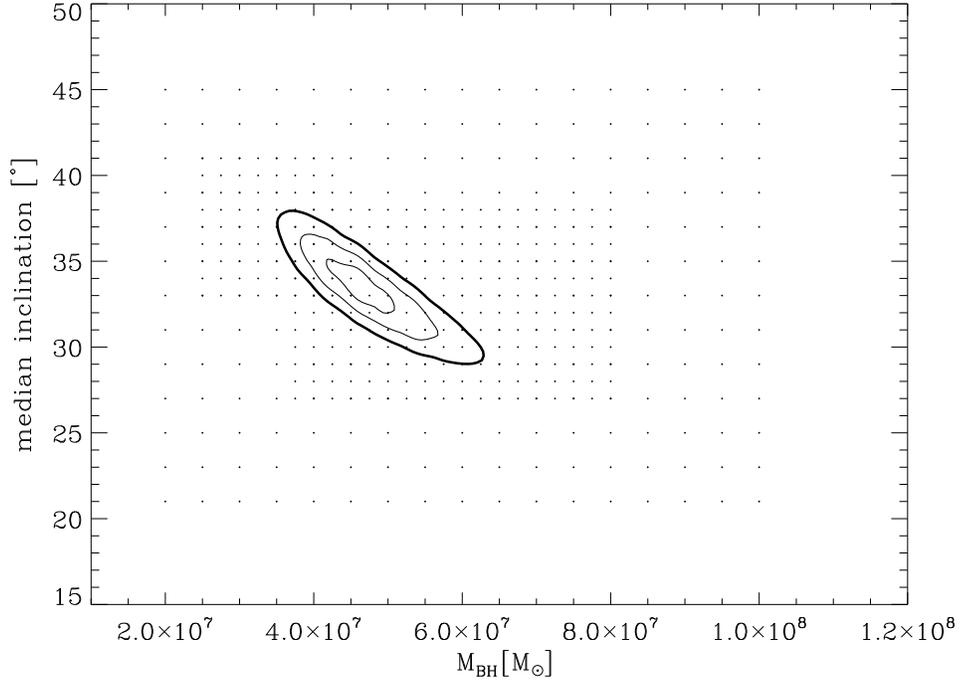}
 \caption{Constraining the mass of the central black hole: The Figure indicates the grid of models 
 (in black hole mass, M$_{\rm{BH}}$, and disk inclination) that was calculated, and the contours show $\Delta$\c2\, 
 in the vicinity of the best fit dynamical models for matching the \h2, kinematics. The minimum \c2\, model is at a 
 $\rm{M_{BH}\sim 4.5 \times10^7 M_{\odot}}$ and a median disk inclination of 34$\degr$. 
 The contours indicate the 1$\sigma$, 2$\sigma$, and 3$\sigma$ confidence levels, respectively (see text for details)}
\label{chi2}
\end{figure*}

\begin{figure*}
\epsscale{0.9}
\plotone{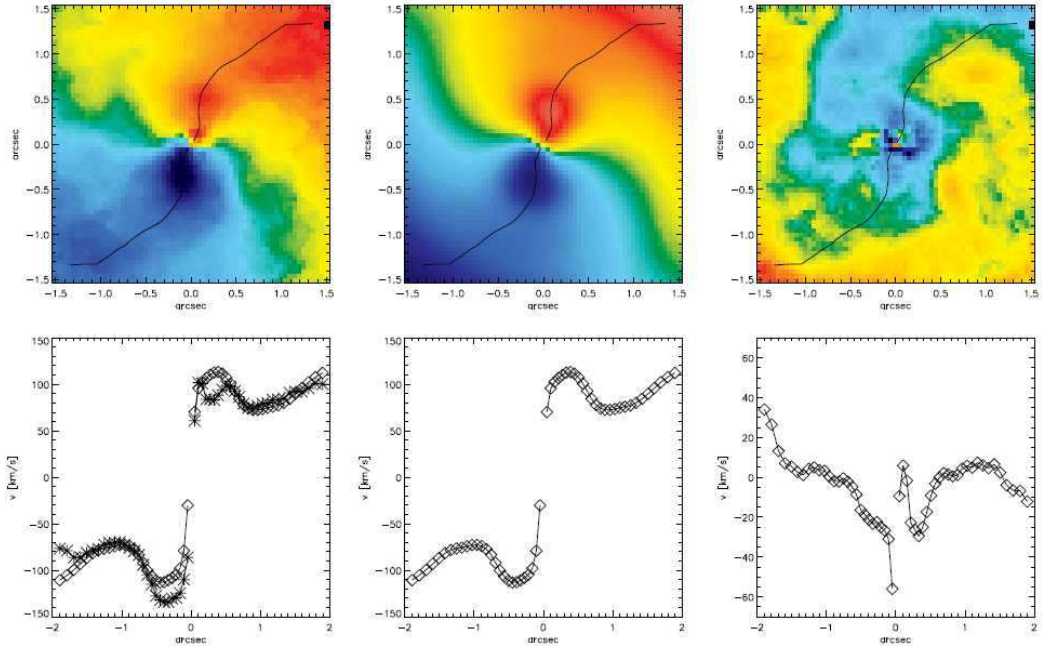}
 \caption{Velocity field of the best-fitting dynamical model (center, top), with a black hole mass of 
 $\rm{4.5 \times10^7 M_{\odot}}$ and a median disk inclination of 34$\degr$, 
 in comparison to the data (top left). The velocity residual 
 (data-model) is shown in the right panel. The velocity curves in the bottom panels
 are extracted along the line-of-nodes (overplotted to the velocity maps), and
 represent the peak velocity curves. The diamonds correspond to the model velocity curve 
 while the crosses correspond to the data. A mismatch in the data and model most prominent 
 beyond r$\lesssim-1\farcs7$ is likely due to the fact that the inclination angle of the modeled 
 gas disk is not well represented in the outermost parts. This has no impact on the derived black hole mass}
\label{best_model}
\end{figure*}

We calculate a grid of possible models for varying disk inclination and central black hole mass 
to get the set of values that best match the observed data.
The best-fitting black hole mass in our tilted-ring model to the
\h2\, kinematics is $\rm{ M_{BH}=4.5^{+1.7}_{-1.0} \times10^7
M_{\odot}}$ for a median inclination of $\sim 34\degr \pm 4\degr$ (error bars 
are given at the 3$\sigma$ level). The best model has a \c2$_{\rm{min}}$ of 8.2. 
This represents the minimum in the $\Delta$\c2\, distribution, shown in Figure~\ref{chi2}. The points represent models. 
The contours were determined by a two-dimensional smoothing spline interpolated 
from these models and represent $\Delta$\c2\, values of 1.0, 4.0, and 9.0. 
This corresponds to 68.3\%, 95.4\%, and 99.7\% confidence levels for 1 degree of freedom, or
1$\sigma$, 2$\sigma$, and 3$\sigma$ confidence levels, respectively.
The associated best-fit model velocity maps are shown in comparison to the data in Figure~\ref{best_model}.
If we would keep the inclination angle fixed to the mean value of 45$\degr$ given by kinemetry the model is not able to reproduce the overall velocity field. The main deviation between model and data appears for radii larger than $\sim0\farcs7$, i.e. outside the radius of influence of the black hole, where the model results in a rotational velocity that is significantly higher than the observed one. The \c2$_{\rm{min}}$ for this  ``best-fit'' model is 16.8 with a best-fit $\rm{ M_{BH}\sim3.0 \times10^7 M_{\odot}}$. This model is thus ruled out to over 3$\sigma$. The entire velocity field is better reproduced by a disk at lower inclination (i.e. more face-on) plus a higher black hole mass that makes up for the decrease in velocity inside the radius of influence of the black hole.

\begin{figure*}
\epsscale{0.9}
\plotone{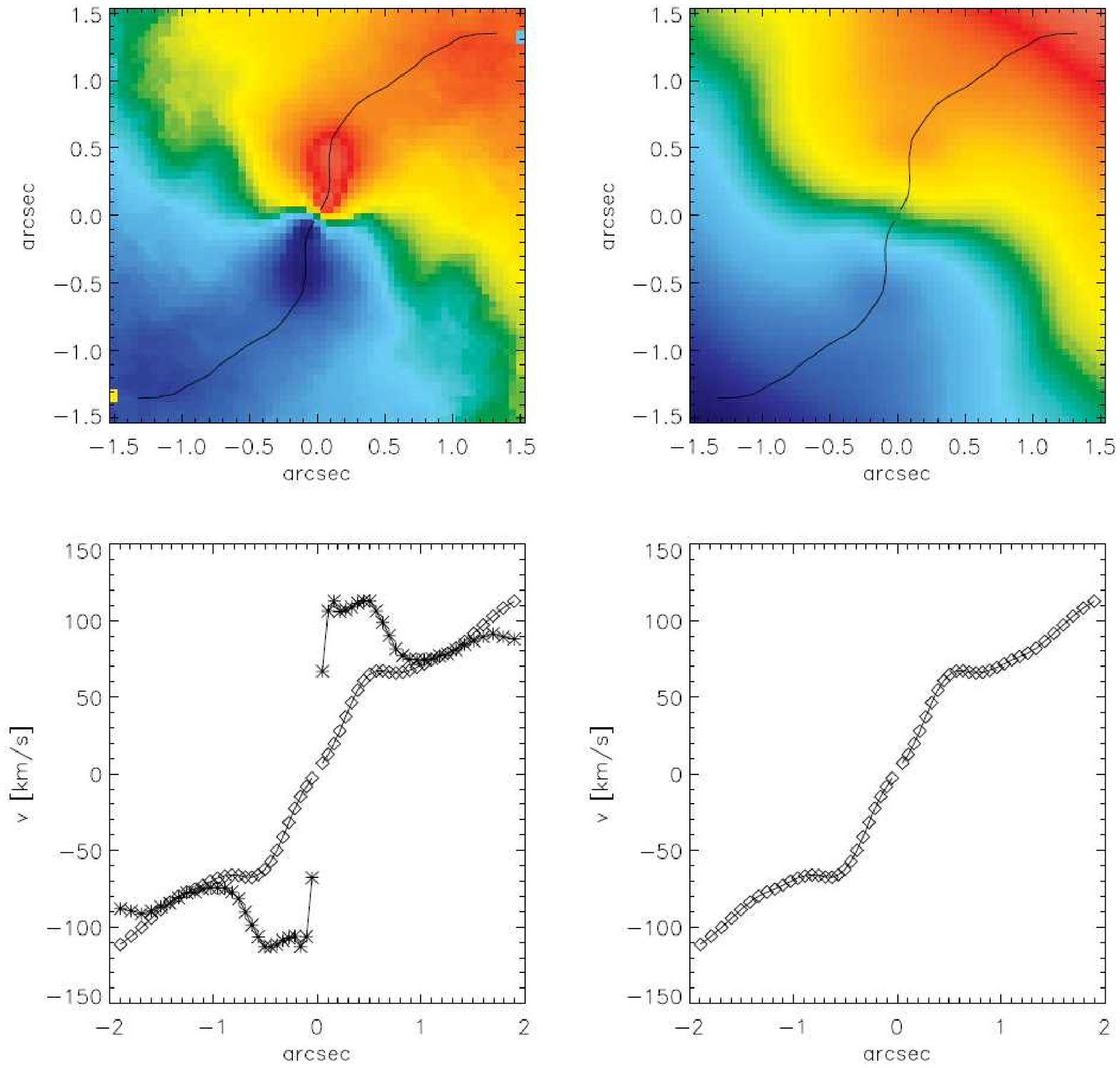}
 \caption{Kinematic evidence for a central black hole: The top panels compare the observed symmetrized
 \h2 velocity field (left) to a velocity field model for the case of zero black
hole mass with only a stellar potential, derived by HN+06 (right). The bottom panels show the velocity
curves extracted along the line-of-nodes (indicated in the top panels). Obviously, this model is not a good fit to the data 
in the central $0\farcs8 \times 0\farcs8$. Only beyond $1\farcs0$ are the model velocity curve (diamonds) in reasonable agreement with the data (crosses)}
\label{no_bh_model}
\end{figure*}

 Figure~\ref{no_bh_model} shows a comparison of the model and the data 
for the case of no central point mass. Here, the gravitational potential is made
up only by the stars. The mass-to-light ratio is 0.72\,$\rm{M/L_{\odot, K}}$ (HN+06). It
is
obvious that this is no good fit in the central $0\farcs5 \times 0\farcs5$. The
modeled rotation only catches up with the data outside $\sim 1\farcs0$, where
the stars clearly dominate the gravitational potential.
The case for no black hole is excluded to very high significance (over 8$\sigma$).

 There are various factors that influence the black hole mass estimate in
our dynamical model and we have done a substantial number of tests
to scrutinize their impact on the best-fit result.
As mentioned in Section~\ref{h2_structure} the assumed
geometry of the disk (warped vs. flat) has a small influence on the black hole
mass. The same holds true for the parameterization of the surface brightness. 
We modeled the kinematics for three different parameterizations of the disk's 
surface brightness profile, with all three being a reasonable fit to the data. 
We found that the black hole mass does change by less than 3\% depending on 
the assumed surface brightness profile of the inner gas disk.
This result is in agreement with the detailed analysis of \cite{marconi06}.\\

Obviously, also the contribution of the stellar potential to the total
gravitational potential influences the resulting best-fit black hole mass. We 
used the two extreme values 0.72\,$\rm{M/L_{\odot, K}}$ and 0.53\,$\rm{M/L_{\odot, K}}$ derived 
by \cite{silge05} through stellar dynamical models at inclination angles 90$\degr$ and 45$\degr$, 
respectively (the value for 20$\degr$ is 0.68\,$\rm{M/L_{\odot, K}}$ and lies in between these two). 
Using 0.53\,$\rm{M/L_{\odot, K}}$, the best-fitting black hole mass increases by 
$\sim8\%$ compared to the best-fit value of $\rm{ M_{BH}=4.5 \times10^7 M_{\odot}}$
(for 0.72\,$\rm\,{M/L_{\odot, K}}$). The uncertainty introduced by the inclination angle
is much larger than this, which is reflected by the fairly large uncertainty limits on our black hole mass measurement.

%-------------- Section 5.4 -----------------------
 
\subsection{Asymmetries in the \h2 velocity field}

The overall shape of the \h2\, velocity field appears to be point-symmetric about
the position of the AGN (the peak in the K-band continuum which 
is coincident with the peak in the \h2 surface brightness). This symmetry is
proven and illustrated by the smooth kinemetry fit.

 However, taking a closer look at the peak velocities in the field, it is
striking, that the peak velocity in the blue dip at $(x,y)=(-0\farcs2, -0\farcs5)$
exceeds the one in the corresponding red peak ($(x,y)=(0\farcs2, 0\farcs5)$)
by $\sim 30$~km~s$^{-1}$ (corresponding to $\sim30\%$
of the peak velocity).

 The reason for this asymmetry is not merely an overall velocity shift, since that
would affect the whole field. This can be ruled out by the fact that the
absolute values of the velocities are in excellent agreement for radii larger
than $\sim 0\farcs8$.

\begin{figure}
\plotone{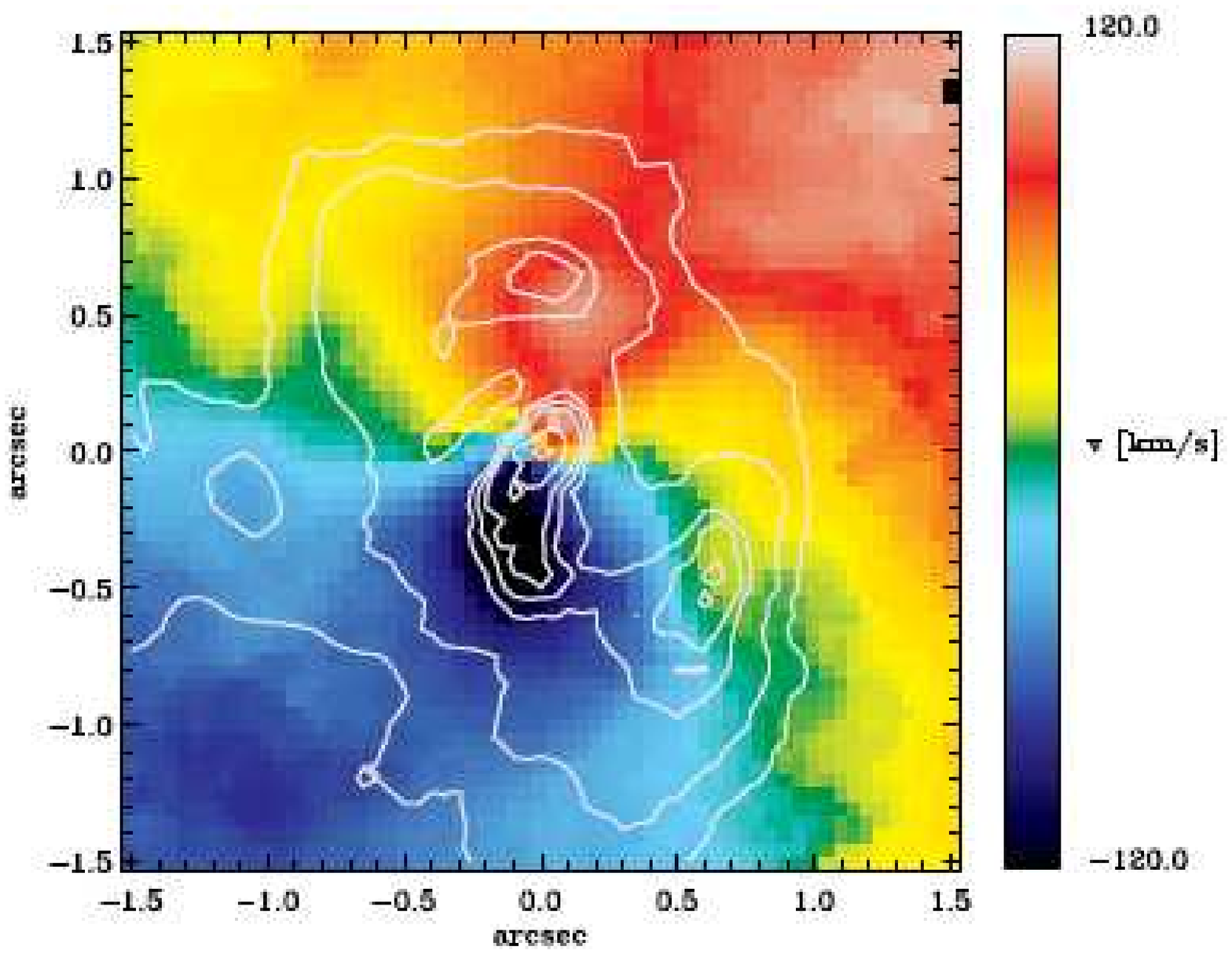}
 \caption{The \h2 surface brightness contours are overplotted to the \h2 velocity map.
The velocity field is asymmetric in the sense that the blue dip is of larger amplitude 
(in the center-of-mass system) than the red peak. This asymmetry is also seen in the flux contours, but such
asymmetries are not captured by our model}
\label{asymmetry}
\end{figure}

 It is instructive to overplot the iso-flux contours to the velocity field (see
Figure~\ref{asymmetry}), since one clearly sees that the asymmetry in the
velocity field is also present in the flux distribution. This leads us to
speculate that the gas density is not the same throughout the disk, as a higher
density leads to a higher flux.  
The density might be increased by gas that is streaming towards the central disk. This might
be of similar origin as the non-rotational motions observed in \fe, \bg, and \si\, but is
not as obvious in molecular hydrogen. The asymmetry is also present in the \bg\,
and \si\, velocity maps, being strongest in \si. What we observe here, might be the
fueling process of the nuclear disk.

\begin{figure*}
\plotone{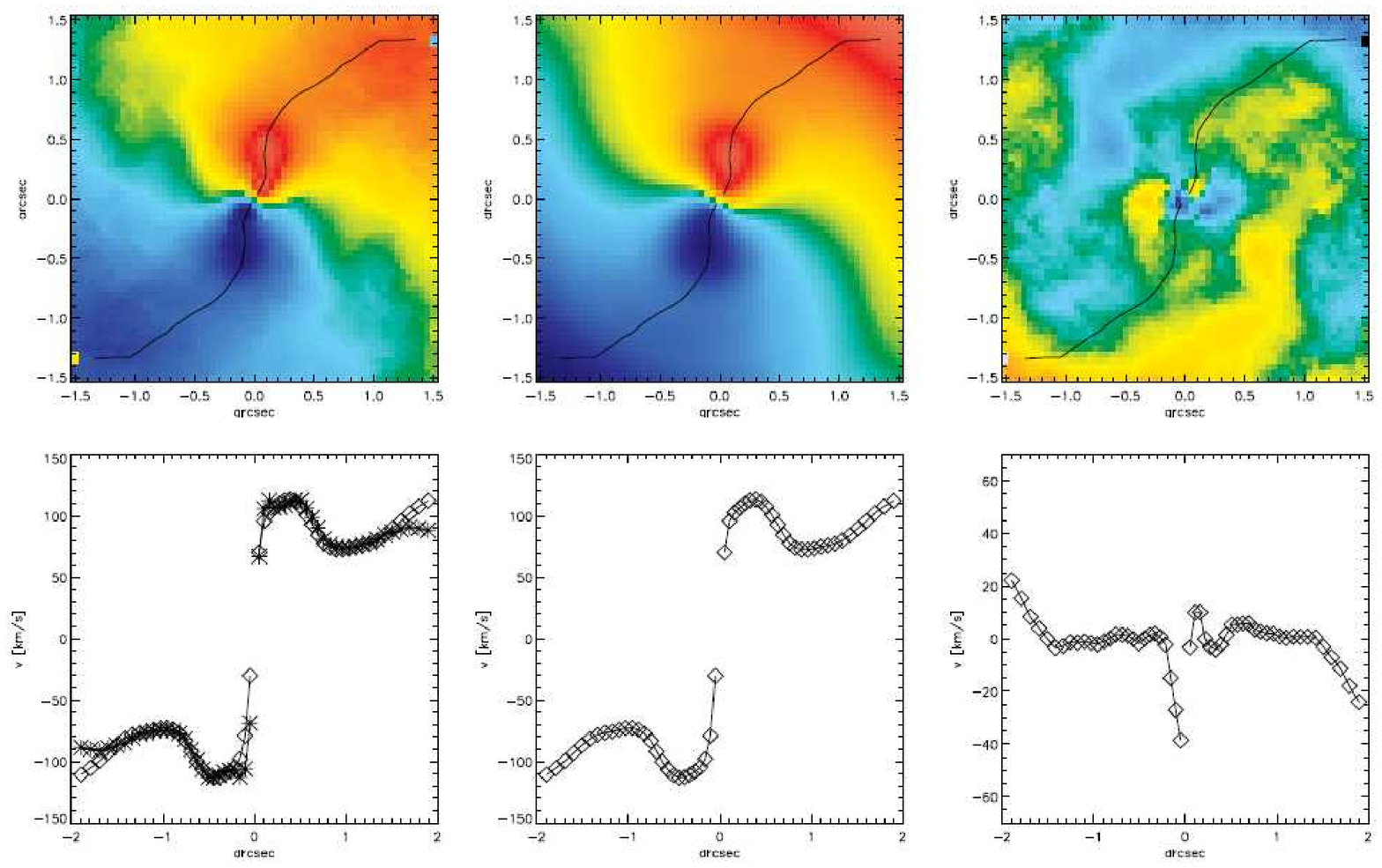}
 \caption{The symmetrized velocity map (top left) in comparison to the model ($M_{BH} = \rm{4.5
 \times10^7 M_{\odot}}$ and $<i> =34\degr$ - top middle panel).
 The residual (data-model) is shown in the right panel. The velocity curves in the bottom  panels
 are extracted along the line-of-nodes (overplotted to the velocity maps), and therefore
 represent the peak velocity curves. The diamonds correspond to the model velocity curve while 
 the crosses correspond to the data. The agreement is very good along the line-of-nodes. The mismatch 
 in data and model for radii beyond r$\sim \pm 1\farcs7$ is likely due to the fact that the inclination 
 angle of the modeled gas disk is not well represented in the outermost parts. This has no impact on the derived black hole mass}
\label{symmetrized}
\end{figure*}

 Our gas dynamical model is by construction symmetric, and therefore not able to
reproduce asymmetries in the velocity structure. For determining the best-fit
black hole mass, we have minimized the difference of the symmetric model to the
asymmetric data, looking for the best compromise. After having gained the
best-fit model (to the original data) we are free to compare this to a
symmetrized version of the \h2 velocity field, $v_{\rm{sym}}(x,y) =
(v(x,y)-v(-x,-y))/2$. The agreement, as shown in Figure~\ref{symmetrized}, is
astonishing.

%-------------- Section 6 -----------------------

\section{Discussion}

The high resolution 2D data have now reached a quality that the existence of a central black hole (or compact mass)
can be demonstrated, and its mass be estimated, even without a sophisticated model, e.g. by just using the concept of the radius of influence of the black hole, r$_{\rm{BH}}$.
This is the region where the mass of the enclosed stars equals the mass of the supermassive black hole:
\begin{equation}
\rm{r_{BH} = \frac{G M_{BH}}{\sigma^2_*}},
\end{equation}
where $\rm{M_{BH}}$ is the mass of the black hole, G is the
gravitational constant and $\sigma_*$ is the velocity dispersion of the stellar spheroid.
In the velocity field, this is the point of minimum rotation, where the black hole stops
to dominate the gravitational potential, before the stars (with their rising rotation curve) 
take over. For our \h2\, velocity this is at
$\sim 0\farcs8$ as seen in Figure~\ref{symmetrized} (left panel). For Cen~A we have the following numbers:
$\sigma_*=138$~km~s$^{-1}$ \citep{silge05}, at D=3.5~Mpc, $\rm{r_{BH}} = 0\farcs8 \pm 0\farcs1 = $ 13.2~pc$\pm1.7$~pc, and therefore we get $\rm{M_{BH} = (6.0 \pm 0.7) \times10^7 \,M_{\odot}}$.

 The observed radius of minimum rotation is independent of the inclination and
so this simple concept provides a nice check on the black hole mass derived via
dynamical modeling. Given the excellent spatial resolution of our AO-assisted
data (FWHM$_{\rm{core}}=0\farcs12$ and FWHM$_{halo}=0\farcs30$) the observed
radius of influence of the black hole is well beyond the radius where PSF
effects start significantly affecting the derived velocity curve.\\

The best-fit black hole mass derived through modeling of the \h2 kinematics,
$\rm{M_{BH}=(4.5^{+1.7}_{-1.0}) \times10^7 M_{\odot}}$ at $i=34\degr \pm 4\degr$, is in good
agreement with the mass derived by HN+06 ($\rm{(6.1^{+0.6}_{-0.8}) \times 10^7
M_{\odot}}$ at $i=45\degr$) using high spatial resolution kinematics of \fe\,
derived from AO-assisted NaCo long slit data (FWHM$=0\farcs11$). The dynamical
model they used is in principle identical to the one described
above, except the fact that we cover the velocity field in two
dimensions and they modeled only four slit positions. Moreover, they did not include the disk 
inclination angle as a free parameter, which is the main reason for their smaller error bars.
Concerning the disk geometry, HN+06 excluded
disk inclination angles below $45\degr$ due to the jet inclination derived by
\cite{tingay98}, who give $50\degr < i < 80\degr$. However, following the analysis of
\cite{hardcastle03}, who use jet-counterjet ratios and apparent motions in the jet,
the jet inclination is most likely in the range $20\degr < i < 50\degr$. HN+06
state that if they were to allow an inclination angle of $i=25\degr$ with respect
to the line-of-sight, their best-fit black hole mass is $\sim 1.2\times10^8 M_{\odot}$.
This is significantly larger than the value we measure. But one has to take into account,
that the modeled gas species are not the same for the two studies. While the molecular
gas has a well-ordered rotation field,
the 2D velocity field of \fe\, (Fig.~\ref{fe}) clearly exhibits two components:
inflow and rotation. This superposition could not be seen in the long-slit data
and was not accounted for in the model of HN+06. They modeled the total \fe\,
kinematics under the assumption of gas rotating in a flat, thin disk.\\

%-------------- Section 6.1 -----------------------

\subsection{Comparison to previous gas dynamical models}

The agreement of our modeling results with the recent analysis of \cite{krajnovic06}
is comfortable: they modeled integral-field Pa$\beta$ kinematics and derived
$\rm{M_{BH}=8.25^{+2.25}_{-4.25} \times10^7 M_{\odot}}$ at $i=25\degr$ 
(where the error bars are also at the 3$\sigma$ confidence level). The resolution
of their seeing limited data is $0\farcs5-0\farcs6$, i.e. a factor of $\sim 4-5$ larger
than ours, but good enough to resolve the radius of influence of the black hole in Cen~A. 
Again, the somewhat larger black hole mass value might be attributed to the fact that \citep{krajnovic06} 
model the kinematics of the ionized gas species Pa$\beta$, which might also be affected by non-gravitational
motions.

\begin{figure}
\plotone{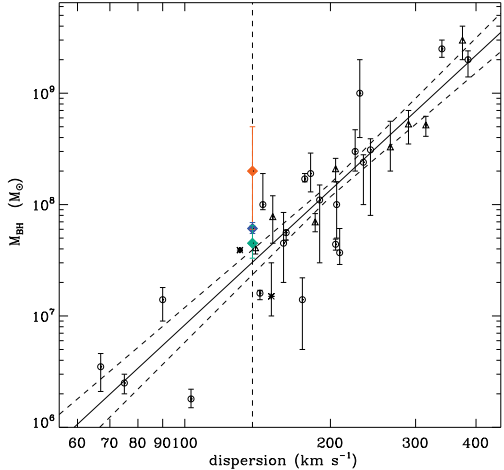}
\caption{Black hole mass (M$_{\rm{BH}}$) vs. stellar velocity dispersion along with the best fit correlation as compiled by \cite{tremaine02}. Mass measurements based on stellar kinematics are denoted by circles, on gas kinematics by triangles, and on maser kinematics by asterisks. Filled diamonds denote mass measurements of Cen~A's black hole by \cite{marconi01} (red), HN+06 (blue), and this work (green). The dashed lines show the 1$\sigma$ confidence levels on the best-fit correlation. Our high resolution two-dimensional kinematic maps and dynamical modeling bring Cen~A in full agreement with the M$_{\rm{BH}}$-$\rm{\sigma}$ relation}
\label{M-sigma}
\end{figure}

 This decrease in black hole mass compared to recent dynamical measurements of ionized gas by
\cite{marconi01} (M$_{\rm{BH}} = 2.0^{+3.0}_{-1.4}\times 10^{8}\rm{M}_{\odot}$),
\cite{neumayer06} (M$_{\rm{BH}} = 6.1^{+0.6}_{-0.8}\times 10^{7}\rm{M}_{\odot}$),
\cite{marconi06} (M$_{\rm{BH}} = 6.5^{+0.7}_{-0.7}\times 10^{7}\rm{M}_{\odot}$), and
\cite{krajnovic06} ($\rm{M_{BH}=8.25^{+2.25}_{-4.25} \times10^7 M_{\odot}}$) brings 
Centaurus~A in agreement with the M$_{\rm{BH}}$-$\rm{\sigma}$ relation \citep{ferrarese00, gebhardt00} (see Fig.~\ref{M-sigma}).
All previous gas dynamical studies were based on ionized gas species, which are most probably all influenced by Cen~A's jet, and are no good tracers of the gravitational potential in the innermost arcsecond. One of the main advantages of our high resolution two-dimensional data is to be able to reveal the jet's influence and to choose the appropriate gas species for the dynamical model.

%-------------- Section 6.2 -----------------------

\subsection{Comparison to stellar dynamical models}

Using stellar kinematics (v, $\sigma$, $h_3$, and $h_4$) derived from Gemini GNIRS data along two slit-positions with seeing-limited resolution of $0\farcs4$ to $0\farcs65$
\cite{silge05} derive a best-fit black hole mass of $2.4^{+0.3}_{-0.2} \times 10^8 M_{\odot}$ for an edge-on model, $1.8^{+0.4}_{-0.4} \times 10^{8}M_{\odot}$ for a model with i=45$\degr$, and $1.5^{+0.3}_{-0.2} \times 10^8 M_{\odot}$ for a model with an inclination of 20$\degr$ (note that theses error bars are at the 1$\sigma$ confidence level). 

Using integral-field stellar kinematics (v, $\sigma$, $h_3$, and $h_4$) derived from the same SINFONI data set used for the underlying gas dynamical study, Cappellari et al. (in prep.) get a best-fit black hole mass of $5 \pm 3 \times 10^7 M_{\odot}$ (3$\sigma$ error bars). This value is more than a factor of $\sim 3$ lower than the value derived by \cite{silge05}, it is, however, in full agreement with the gas dynamical measurement presented in the current paper.

 This is an important result for the general comparison of stellar and gas
kinematical modeling of black hole masses. Up to now only a couple of these
comparative cases have been investigated and in half of the cases with negative
results \citep[e.g.][]{cappellari1459, shapiro06}.

 This is the first case where the modeling techniques can be tested on stellar
and gas kinematical data extracted from the same complete data set. The fact
that we deal with integral-field kinematics allows the constraints on the black
hole mass to be significantly tightened compared to long-slit observations. The
excellent agreement between gas and stellar dynamical modeling results thus
provides stringent evidence for the reliability of both techniques.

%-------------- Section 7 -----------------------

\section{Conclusions}

This work presents observations of the nearby active elliptical galaxy NGC~5128 (Cen~A)
with the adaptive-optics assisted integral field spectrograph SINFONI at the VLT. Our
K-band data used to measure the black hole mass in Cen~A from emission line kinematics have a spatial resolution
of $0\farcs12$ and an estimated Strehl ratio of $\sim 17\%$. The field of view is $3\arcsec \times 3\arcsec$.\\
In our H- and K-band data we detect the following emission lines: \fe, \si, \he, \bg, \ca,
and several vibrational transitions of molecular hydrogen, \h2\, (the strongest is 1-0\,S(1)
at 2.12$\mu$m). The main results of our analysis are:

\begin{itemize}

\item Analysing the velocity fields of \si, \bg, \fe, and the strongest \h2\, line, we find that
the surface brightness and also the motion of the gas species is increasingly influenced
(or produced) by the jet when going from low to high excitation lines.
The velocity fields of \si, \bg, and \fe\, clearly exhibit two components: 1) rotational
motion at a major angle of $\sim 150\degr$, consistent with an orthogonal disk-jet picture, and
2) non-rotational motion along the direction of the jet, that is consistent with a back-flow
of gas along the side of the jet's cocoon. This non-rotational component is strongest for \si.

\item The surface brightness of \h2\, shows an indication of a central gas disk at a position angle of
P.A.$\sim 136\degr$ and a minor-to-major axis ratio of 0.67. The bright spots NE and SW of the
nuclear disk remind of the lobes seen on much larger scales in jet-gas interactions.

\item The overall velocity field of \h2\, shows beautiful point-symmetry about the unresolved nucleus.
However, the \h2\, peak velocities are asymmetric inside the central $0\farcs7$. This asymmetry is also seen
in the surface brightness map of \h2, and might be explained by denser gas that leads to an increase
in flux. The density could be increased by gas that is streaming towards the
central disk. This might be similar to the non-rotational motions observed in \fe, \bg, and \si\, but is
not as obvious in molecular hydrogen. It is possible that we see the fueling process of the nuclear
disk.\\
Another influence of the jet on the velocity field might be the twists in the zero velocity curve (at the positions where the
lobes appear in the surface brightness map of \h2).
The rotational major axis (line-of-nodes) runs from SE to NW, with the SE side blue- and the
NW side redshifted with respect to the nucleus. It follows the shape of an ``S". The velocity
field is reminiscent of a warped disk.
Judging from the smooth velocity field of the \h2\, gas, the major part of molecular hydrogen is well settled in the total gravitational potential $\Phi_{\star} + \Phi_{\rm{BH}}$, and is therefore a good tracer of the central black hole mass.

\item Under the assumption of gas moving on circular orbits, we fit the geometry of the gas disk
with a set of tilted rings, using kinemetry \citep{davor06}. The mean P.A. is 155$\degr$ with
twists of $\pm14\degr$. The median inclination angle is $\sim45 \degr \pm 12 \degr$.

\item We construct a tilted-ring model for the central \h2\, gas motions, in the combined potential
of the stars and the black hole. We account for the velocity dispersion via a pressure term in the
isotropic, axisymmetric Jeans' equation. The geometry
of the disk is fixed via the kinemetry fit to the velocity field, but we allow the
overall inclination angle to vary. Our best-fit model has a mean inclination angle of $34\degr
\pm 4\degr$, and a black hole mass of $\rm{M_{BH}=(4.5^{+1.7}_{-1.0}) \times10^7 M_{\odot}}$. We find that
the inclination angle, derived through the kinemetry fit is not able to reproduce the overall
velocity field. We conclude that the optimal choice would be to simultaneously fit the disk geometry and
the black hole mass.

 As our dynamical model is by construction symmetric about the center, it fails
to reproduce the asymmetry in the \h2\, data. When symmetrising the velocity field,
the agreement of data and model is excellent.

\item Our black hole mass measurement is somewhat smaller but still consistent with previous gas measurements using ionized gas species. It is substantially smaller than the stellar dynamical model of \cite{silge05}. However, it is in perfect agreement with the stellar dynamical measurement of Cappellari et al. (in prep.) who extract the stellar kinematics from the same SINFONI data set.

\item Our black hole mass determination brings Cen~A in agreement with the $\rm{M_{BH} - \sigma}$ relation \citep{ferrarese00, gebhardt00}.

\end{itemize}

\section*{Acknowledgments}
We thank the Paranal
Observatory Team for the support during the observations. We are grateful to
Davor Krajnovi\'c for making the Kinemetry software publicly available. We thank Martin Hardcastle 
for kindly providing unpublished VLA data.
This work has been made possible through support by the Netherlands Research School for Astronomy NOVA.
NN acknowledges support from the Christiane-N\"usslein-Volhard Foundation.
MC acknowledges support from a PPARC Advanced Fellowship (PP/D005574/1).

\label{lastpage}

\end{document}